\documentclass{article}

\usepackage{amsmath}
\usepackage[pdftex]{graphicx}
\usepackage{enumerate}
\usepackage{natbib}
\usepackage{url} 

\usepackage{amssymb}
\usepackage{hyperref}
\usepackage{subcaption}

\usepackage{siunitx}
\RequirePackage{tikz}
\usetikzlibrary{arrows}
\usetikzlibrary{fit}
\usetikzlibrary{shapes.geometric}

\newcommand{\appref}[1]{\hyperref[#1]{Appendix~\ref*{#1}}}

\newcommand{\bX}{\mathbf{X}}              
\newcommand{\bY}{\mathbf{Y}}              
\newcommand{\bmu}{\boldsymbol{\mu}} 
\newcommand{\bSigma}{\boldsymbol{\Sigma}}     
\newcommand{\bA}{\mathbf{A}}              
\newcommand{\bI}{\mathbf{I}}              
\newcommand{\bD}{\mathbf{D}}              

\newcommand{\bLambda}{\boldsymbol{\Lambda}}

\DeclareMathOperator{\E}{E}              
\DeclareMathOperator{\var}{var}          
\DeclareMathOperator{\cov}{cov}          
\DeclareMathOperator{\corr}{Cor}        
\DeclareMathOperator{\lik}{\mathit{L}}   
\DeclareMathOperator{\loglik}{\ell}      

\DeclareMathOperator{\N}{\mathit{N}}      
\DeclareMathOperator{\Invgamma}{\mathit{Inv-gamma}} 
\DeclareMathOperator{\Exp}{\mathit{Exp}}      

\newcommand{\Ep}[1]{\E \left[ #1 \right]}                 
\newcommand{\varp}[1]{\var \left( #1 \right)}             
\newcommand{\covp}[1]{\cov \left( #1 \right)}             
\newcommand{\corrp}[1]{\corr \left( #1 \right)}           

\newcommand{\Np }[2]{\N  \left( #1,#2 \right)}     
\newcommand{\Invgammap}[2]{\Invgamma \left( #1,#2\right)}     
\newcommand{\Expp}[1]{\Exp \left( #1 \right)} 

\newcommand{\sigmaz}{\sigma_Z}
\newcommand{\sigmah}{\sigma_H}
\newcommand{\sigmaf}{\sigma_{F \mid H}}
\newcommand{\sigmahm}{\sigma_m}
\newcommand{\sigmafm}{\varphi_m}
\newcommand{\sigmaha}{\sigma_a}
\newcommand{\sigmafa}{\varphi_a}
\newcommand{\esigmahm}{\psi}
\newcommand{\esigmafm}{\theta}
\newcommand{\sigmadh}{\sigma_{\Delta_H}}
\newcommand{\sigmadf}{\sigma_{\Delta_{F \mid H}}}
\newcommand{\sigmadw}{\sigma_{\Delta_W}}

\begin{document}

\title{\bf On constraining projections of future climate using observations and simulations from multiple climate models}

\author{Philip G. Sansom, David B. Stephenson \\
        University of Exeter \\
        and \\
        Thomas J. Bracegirdle \\
        British Antarctic Survey}

\maketitle

\begin{abstract}
Numerical climate models are used to project future climate change due to both anthropogenic and natural causes.
Differences between projections from different climate models are a major source of uncertainty about future climate.
Emergent relationships shared by multiple climate models have the potential to constrain our uncertainty when combined with historical observations.
We combine projections from 13 climate models with observational data to quantify the impact of emergent relationships on projections of future warming in the Arctic at the end of the 21st century.
We propose a hierarchical Bayesian framework based on a coexchangeable representation of the relationship between climate models and the Earth system.
We show how emergent constraints fit into the coexchangeable representation, and extend it to account for internal variability simulated by the models and natural variability in the Earth system.
Our analysis shows that projected warming in some regions of the Arctic may be more than \SI{2}{\celsius} lower and our uncertainty reduced by up to \SI{30}{\percent} when constrained by historical observations.
A detailed theoretical comparison with existing multi-model projection frameworks is also provided.
In particular, we show that projections may be biased if we do not account for internal variability in climate model predictions.
\end{abstract}

\vfill

\newpage

\section{Introduction}
\label{sec:intro}

Scientific inquiry into complex systems such as the climate naturally leads to multiple models of a system.
The situation in climate science is unusual since different climate models are not treated as incompatible or competing.
Instead, each model is treated as a plausible representation of the climate system \citep{Parker2006}.
This has led to the use of multi-model ensembles to quantify the uncertainty in projections of future climate introduced by choices in model design, usually referred to simply as model uncertainty \citep{Tebaldi2007}.
Statistical methods are required to interpret projections from multi-model ensembles and to make credible probabilistic inferences about future climate change.

In addition to model uncertainty, projections of future climate are subject to a number of other sources of uncertainty. 
Model inadequacy refers to differences between the models and the Earth system, e.g., missing processes \citep{Craig2001,Stainforth2007}.
We intuitively think of climate as the distribution of weather. \citep{Stainforth2007,Stephenson2012,Rougier2014}.
Natural variability refers to the range of possible conditions we might experience and is sometimes referred to as sampling uncertainty, since we only observe a single actualization of the Earth system \citep{Chandler2013}.
Climate models attempt to simulate natural variability by performing multiple simulations from slightly different initial conditions.
This is known as internal variability or initial condition uncertainty.
Climate projections are also subject to forcing uncertainty and parameter uncertainty.
Forcing uncertainty arises due to uncertainty about future emissions of greenhouse gases, both anthropogenic and natural.
Parameter uncertainty refers to uncertainty about choice of the internal parameters in climate models \citep{Collins2007}.
Forcing uncertainty is usually circumvented by making projections rather than predictions of future climate, i.e., predictions conditioned on an assumed future emissions scenario, e.g., \citet{Moss2010}.
The computational cost of running sufficiently large perturbed-parameter experiments to span the full range of parameter uncertainty for a single climate model can be prohibitive.
Therefore multi-model ensembles usually consist of a set of ``best estimates'', i.e., a single version of each model with the internal parameters fixed \citep{Knutti2010}.

In this paper, we develop a hierarchical Bayesian framework for combining projections from multiple models, applied to projecting climate change in the Arctic at the end of the 21st century.
The proposed framework separates model uncertainty and model inadequacy, and accounts for internal variability and natural variability in future projections.
In addition, we are able constrain projections of future climate using historical observations (where suitable constraints have been identified) while accounting for uncertainty in the observations.

In order to make projections of future climate from multi-model ensembles, it is necessary to make assumptions about the relationship between climate models and the Earth system.
One widely used assumption is that skill in reproducing past climate implies skill in projecting future climate.
Climate scientists have long recognized that no single model will perform best for all variables or in all regions \citep{Lambert2001,Jun2008}.
Various approaches have been proposed for weighting projections from multiple climate models based on their ability to reproduce past climate, these include heuristics \citep{Sanderson2015b,Sanderson2015a,Knutti2017}, multiple regression \citep{Greene2006,Bishop2013}, pattern scaling \citep{Shiogama2011,Watterson2011} and Bayesian Model Averaging \citep{Min2006,Bhat2011}.
However, \citet{Weigel2010} demonstrated that weights that do not accurately reflect the projection skill of the models can lead to less reliable projections than weighting all models equally.
Long-term climate projection involves extrapolation to states that have not been observed in recent Earth history.
Therefore, the ability to reproduce observed data does not guarantee skill for projecting future events \citep{Oreskes1994}.
However, we should certainly be cautious when interpreting projections from models that are not able to adequately reproduce observed data, although how such performance should be quantified remains an open question \citep{Knutti2010}.

Weighting all models equally implies that each climate model performs equally well for simulating future climate \emph{change}.
This has led to the alternative assumption that any bias between the models and the Earth system remains approximately constant over time \citep{Buser2009}.
Under this assumption, two main interpretations of multi-model experiments have emerged \citep{Stephenson2012}.
The ``truth plus error'' approach treats the output of each model as the ``true'' state of the Earth system plus some error that is unique to each model \citep{Cubasch2001,Tebaldi2005,Furrer2007,Furrer2007a,Smith2009,Tebaldi2009}.
The ``exchangeable'' approach treats the Earth system as though it were just another climate model, i.e., our inferences about the future climate of the Earth system should be the same as for a climate model with an identical historical climate \citep{Raisanen2001,Annan2010,Annan2011}.
Neither interpretation is entirely satisfactory.
The truth-plus-error interpretation implies that we can improve the precision (but not necessarily the accuracy) of our projections of future climate simply by adding more models to our ensemble. \citep{Annan2010,Knutti2010b}.
The exchangeable interpretation ignores the inherent differences between computer models and the physical systems they seek to represent \citep{Craig2001,Kennedy2001}.

Both the truth-plus error and exchangeable approaches acknowledge differences between models, and between individual models and the Earth system.
What is missing are differences from the Earth system that are \emph{common} to all models.
All climate models are based on a shared but limited knowledge of the Earth system and face similar technological constraints (e.g., similar numerical methods, available CPU time, memory, etc.), so common limitations will inevitably occur \citep{Stainforth2007}.
To address this issue, \citet{Chandler2013} and \citet{Rougieretal2013} independently introduced the idea of representing common model errors as a discrepancy between the expected state of the Earth system and a ``consensus'' or ``representative'' model.
This has the effect of separating model uncertainty (differences between models) from model inadequacy (common differences between models and the Earth system).

\begin{figure}[t!]
  \center
  \includegraphics[width=0.66\textwidth]{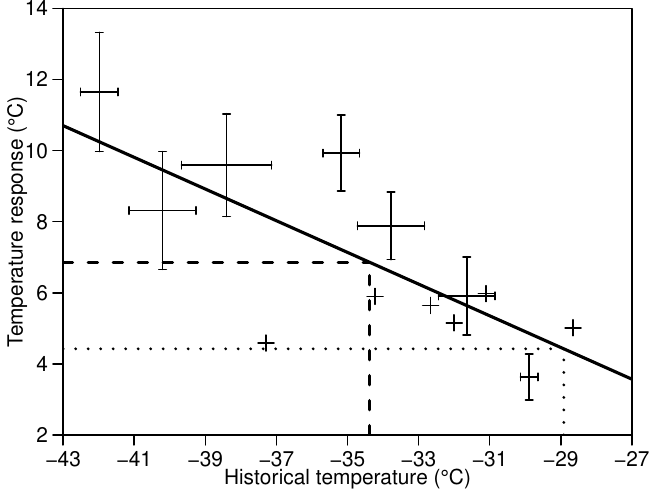}
  \caption{Near-surface warming in the Canadian Arctic Archipelago.
           Thirty-year mean temperature change between 1975--2005 and 2069--2099 as simulated by an ensemble of 13 climate models under the RCP4.5 mid-range mitigation scenario, for a \ang{2.5}$\times$\ang{2.5} grid box centered on Melville Island (\ang{76}N,\ang{111}W).
           Crosses mark the mean climate and climate response simulated by each model.
           Whiskers indicate the range of 30-year mean outcomes from the initial condition runs of each model.
           The dashed line indicates the mean climate and climate response of the ensemble.
           The solid line is a simple linear regression estimate of the emergent relationship between the climate response and the historical climate.
           The dotted line indicates the observed historical climate and projected climate response given the estimated emergent relationship.
          }
  \label{fig:example}
\end{figure}

Historical observations have been used in a variety of ways to constrain projections from individual models \citep{Collins2012}.
However, if systematic relationships existed between the historical states and climate responses simulated by \emph{multiple} models, then it might be possible to constrain projections of future climate in a multi-model ensemble without assigning weights to individual models.
One of the earliest examples of such a relationship was noted by \citet{AllenIngram2002} who referred to it as an ``emergent constraint'', since it emerged from analysis of a collection of model simulations rather than by direct calculation based on theory.
There is now a growing body of evidence that such relationships may exist at the local or process level, and even at the global level \citep{Halletal2019,Brient2020}.
In general, we prefer the term ``emergent relationship''.
We reserve the term ``emergent constraint'' for when physical insight indicates that the relationship should also hold in the Earth system.
\autoref{fig:example} shows an example of a well understood emergent constraint on surface temperature in the Arctic due to albedo feedbacks caused by variations in sea-ice coverage simulated by the models \citep{Bracegirdle2012}.
Other examples of emergent relationships have been found in the cryosphere \citep{HallQu2006,Boe2009}, atmospheric chemistry \citep{Eyring2007,Karpechko2013}, the carbon cycle \citep{Cox2013,Wenzel2014} and various other areas of the Earth system.
The constraint on Equilibrium Climate Sensitivity proposed by \citet{Cox2018} is a rare example that was derived from theory, then found to be present in a collection of model simulations.
Simple linear regression is often used to estimate emergent relationships.
However, projection either implicitly treats the Earth system as exchangeable with the models \citep[e.g.,][]{Bracegirdle2012,Bracegirdle2013}, or simply excludes all models that fall outside the plausible range of the observations \citep[e.g.,][]{HallQu2006,Qu2014}.

Multi-model ensembles are sometimes known as ``ensembles of opportunity'' since models are not systematically selected to span model uncertainty, and cannot be considered a random sample from some larger population \citep{Stephenson2012}.
In particular, several research centers maintain more than one model, and models from different centers often share common components \citep{Knutti2013}.
Similar models are likely to give similar outputs, leading to clustering that could result in biased inferences if not properly accounted for.
This is especially important when analyzing emergent constraints since a large cluster of outlying models could strongly influence any regression relationship.
Therefore, care is required to ensure that our assumptions in representing model uncertainty and inadequacy are satisfied.

Model uncertainty/inadequacy tends to dominate other sources of uncertainty in long-term climate projections \citep{Hawkins2009,Yip2011}. 
However, there is now a significant body of work highlighting the importance of internal variability and natural variability \citep{Deser2012a,Thompson2015,McKinnon2018}.
Several studies have shown that the contribution of internal variability is non-negligible compared to model uncertainty for some variables at the global scale, and particularly at the regional scale \citep{Hawkins2009,Hawkins2011,Northrop2014}.
The internal variability simulated by each model is indicated by the whiskers in \autoref{fig:example}.
Current frameworks for multi-model inference often ignore internal variability and select a single initial condition run from each model \citep[e.g.,][]{Tebaldi2005,Smith2009,Bishop2013} or take the average over all runs from each model \citep[e.g.,][]{Watterson2011,Bracegirdle2012}.
Measurement and representation errors in our observations of the climate system can also contribute significant observation uncertainty.
Some authors have accounted for observation uncertainty \citep[e.g.,][]{Bowman2018,Cox2018}, but it is frequently ignored and plays an important role if we want to constrain future projections using past observations.


The remainder of this study proceeds as follows.
\autoref{sec:data} outlines the data used to project future warming in the Arctic.
In \autoref{sec:framework}, we develop a hierarchical Bayesian framework for inferring time mean future climate from multi-model experiments for any future time period and location for which we have model simulations, conditional on simulations of a recent period for which we have corresponding observations.
We do not attempt to account for spatial correlation between locations or temporal variation within each time period.
\autoref{sec:discussion} compares our proposed framework to existing multi-model ensemble approaches.
In \autoref{sec:results}, we apply our framework to the projection of future climate change in the Arctic.
We end with concluding remarks in \autoref{sec:conclusion}.

\section{Future climate change in the Arctic}
\label{sec:data}

The magnitude of the projected warming in the polar regions is much greater than at lower latitudes \citep{HollandBitz2003}.
We use outputs from 13 climate models participating in the World Climate Research Programme's Coupled Model Intercomparison Project phase 5 \citep[CMIP5,][]{Taylor2012} to investigate the impact of emergent constraints 
on projections of winter (December-January-February) near-surface (\SI{2}{\meter}) temperature change in the Arctic.
We compare the 30 year average winter temperature between two time periods.
The climate models included and the number of initial condition runs available for each time period are listed in \autoref{tab:models}.
The historical period is defined as between December 1975 and January 2005, as simulated under the CMIP5 historical emissions scenario.
The future period of interest is between December 2069 and January 2099, as simulated under the RCP4.5 mid-range mitigation scenario \citep{Moss2010}.
The domain of interest is \ang{45}N--\ang{90}N, including not only the Arctic Ocean but also the Bering Sea and the Sea of Okhotsk, both of which also currently experience significant seasonal sea ice coverage.
Due to the presence of seasonal ice coverage and the complexity associated with modeling it, both model uncertainty and internal variability in near-surface temperature are much greater in the Arctic than at lower latitudes \citep{Northrop2014}.
Prior to analysis, data from all models were interpolated bicubically to a common grid with equal \ang{2.5} spacing in both longitude and latitude.

\begin{table}[t!]
  \caption{Multi-model ensemble. 
           Number of runs available from each model for the historical and future time periods.}
  \label{tab:models}
  \centering
    \begin{small}
      \begin{tabular}{llcc}
        \hline \hline
        \centering 
        & & \multicolumn{2}{c}{Runs} \\
        \cline{3-4}
        Modeling center & Model & Historical & Future \\
        \cline{3-4}
        & & $N_{Hm}$ & $N_{Fm}$ \\
        \hline
        BCC          & BCC-CSM1.1(m) & 3 & 1 \\
        CCCMA        & CanESM2       & 5 & 5 \\
        NSF-DOE-NCAR & CESM1(CAM5)   & 3 & 3 \\
        ICHEC        & EC-EARTH      & 8 & 9 \\
        LASG-CESS    & FGOALS-g2     & 5 & 1 \\
        NOAA GFDL    & GFDL-ESM2G    & 1 & 1 \\
        NASA GISS    & GISS-E2-R     & 6 & 6 \\
        MOHC         & HadGEM2-ES    & 4 & 4 \\
        INM          & INM-CM4       & 1 & 1 \\
        IPSL         & IPSL-CM5A-MR  & 3 & 1 \\
        MIROC        & MIROC5        & 5 & 3 \\
        MPI-M        & MPI-ESM-LR    & 3 & 3 \\
        MRI          & MRI-CGCM3     & 3 & 1 \\
        \hline
        Total & & 50 & 39 \\    \hline
      \end{tabular}
    \end{small}
\end{table}

Observational data in the Arctic are very sparse and no spatially complete data sets exist that include estimates of observational uncertainty.
Therefore, we combine four contemporary reanalysis data sets (ERA-Interim \citep{Dee2011}, NCEP CFSR \citep{Saha2010}, JRA-25 \citep{Onogi2007}, NASA MERRA \citep{Rienecker2011}) using the methodology proposed in \autoref{sec:inference}.
Reanalysis data was interpolated to the same grid as the models.

\section{A hierarchical framework for multi-model experiments}
\label{sec:framework}



\begin{figure}[t]
  \centering
  \begin{tikzpicture}[->, >=stealth', shorten >=1pt, auto, inner sep=0pt, 
    minimum height=2.54cm, minimum width=2.54cm, node distance=5.08cm,
    thick, every node/.style={font=\sffamily\Large\bfseries,align=center},
    scale=0.5, transform shape] 

    \node[rectangle, draw, fill=gray!30] (M)
      {$\mu_H, \esigmahm^2$ \\ $\mu_F,\beta,\esigmafm^2$} ;
    \node[rectangle, draw, fill=gray!30] (R)  [left of=M] 
      {$\sigmah^2, \nu_H$ \\ $\sigmaf^2, \nu_F$} ;
    \node[rectangle, draw, fill=gray!30] (U)  [right of=M]     
      {$\sigmadh^2, \nu_{Ha}$ \\ 
       $\sigmadf^2, \nu_{Fa}$} ; 
            
    \node[circle, draw, fill=gray!30]    (X)  [above of=R]  
      {$X_{Hm}, \sigmahm^2$ \\ $X_{Fm}, \sigmafm^2$} ;
    \node[circle, draw, fill=gray!30]    (Y)  [above of=U] 
      {$Y_H, \sigmaha^2$ \\ $Y_F, \sigmafa^2$} ;
      
    \node[diamond, draw, , fill=gray!30, minimum height=3.59cm, minimum width=3.59cm]                            (Xr) [above of=X]  
      {$X_{Hmr}$ \\ $X_{Fmr}$} ;
    \node[circle, draw, fill=gray!30]    (Ya) [above of=Y]             
      {$Y_{Ha}$ \\ $Y_{Fa}$} ;
    \node (wea) [above of=Ya, node distance=2.22cm]  
      {Actualized \\ climate} ;

    \node[rectangle, draw, fill=gray!30] (W)  [right of=U] 
      {$\sigmaz^2$ \\} ;
    \node[diamond, draw, fill=gray!30, minimum height=3.59cm, minimum width=3.59cm, node distance=10.16cm]     (Z)  [above of=W] 
      {$Z_H$ \\} ;
      
    \node (mmm) [below of=M,  node distance=2.54cm]  
      {Representative \\ climate} ;
    \node (md)  [below of=R,  node distance=2.54cm]  
      {Model \\ uncertainty} ;
    \node (se)  [below of=U,  node distance=2.54cm]  
      {Model \\ inadequacy}  ;
    \node (oe)  [below of=W,  node distance=2.54cm]  
      {Observation \\ uncertainty} ;
    \node (run) [above of=Xr, node distance=2.22cm]  
      {Model output} ;
    \node (obs) [above of=Z,  node distance=2.22cm]  
      {Observations}  ;

    \node (obs) [above of=R,  node distance=2.86cm]  
      {$m = 1,\ldots,M$}  ;
      
    \node (dat)  [right of=Z, node distance=3.81cm]  
      {Data \\ level} ;
    \node (cli)  [below of=dat]  
      {Latent \\ level} ;
    \node (par)  [below of=cli]  
      {Parameter \\ level} ;

    \path[every node/.style={font=\sffamily\small}]
    (M)  edge (X)
    (M)  edge (Y)
    (R)  edge (X)
    (U)  edge (Y)
    (X)  edge (Xr)
    (Y)  edge (Ya)
    (Ya) edge (Z)
    (W)  edge (Z);
    
    \node [draw,dashed,above of=X, node distance=2.54cm, minimum height=10.16cm, minimum width=5.08cm] {};

  \end{tikzpicture}
  \caption{Graphical representation.
           The proposed framework represented as a directed acyclic graph.
           Diamonds represent data, circles represent latent quantities, and squares represent parameters.
           The dashed box represents the multi-model ensemble.
           The actualized climate $Y_{ta}$ is placed at the data level to emphasize its relationship with the model runs $X_{tmr}$.}
  \label{fig:dag}
\end{figure}
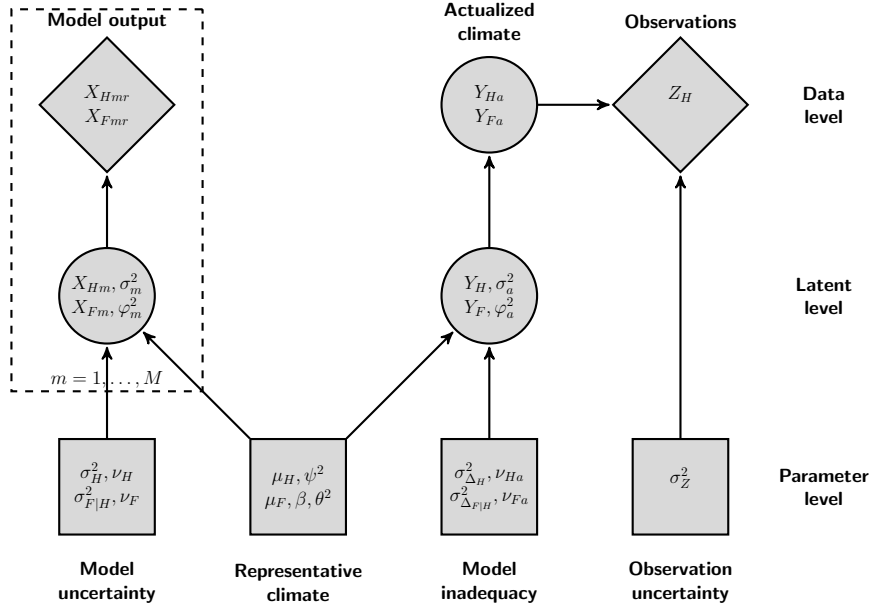

The proposed framework is summarized in graphical form in \autoref{fig:dag}.
We compare one historical time period denoted $H$ and one future time period denoted $F$, conditioned on a single future emissions scenario.
The top level of \autoref{fig:dag} consists of quantities for which we have data, i.e., model outputs ($X_{Hmr}$,$X_{Fmr}$) and observations ($Z_H$).
The mid-level consists of the climates of the individual models and the Earth system, quantified by the means ($X_{Hm}$,$X_{Fm}$,$Y_H$,$Y_F$) and variances ($\sigma_m^2$,$\psi_m^2$,$\sigma_a^2$,$\psi_a^2$) of the simulated or plausible conditions during each time period.
The bottom level consists of parameters quantifying model uncertainty, the ``representative'' climate of the ensemble, model inadequacy, and observation uncertainty.
All of these quantities will be fully defined in the development that follows.

We proceed in three stages.
First, we propose a hierarchical model for the outputs of the multi-model ensemble (left hand side, \autoref{fig:dag}).
Second, we propose a similar hierarchical model for the climate of the Earth system (middle right, \autoref{fig:dag}).
Finally, we specify a model for the relationship between the actualized climate and the observations (right hand side, \autoref{fig:dag}).

\subsection{The multi-model ensemble}
\label{sec:mme}

Suppose we have an ensemble of $M$ climate models.
Each model performs a number of runs of the historical and future time periods, conditioned on a single future emissions scenario.
Each run is initialized from slightly perturbed initial conditions.
Let $X_{tmr}$ be the output of run $r$, during time period $t = \lbrace H,F \rbrace$, by model $m = 1,\ldots,M$.
The outputs $X_{tmr}$ are assumed to be time averages over periods of equal length.
The number of runs of each model for each time period is denoted $R_{tm}$, i.e., $r = 1,\ldots,R_{tm}$ for model $m$ in time period $t$.
We do \emph{not} require that the number of runs from each model be equal, or that the number of runs of each period by a particular model be equal (frequently $R_{Fm} < R_{Hm}$).
Each model is attempting to simulate the same target, i.e., the climate of the Earth system under a specific emissions scenario.
Therefore, we assume \textit{a priori} that the model outputs $X_{tmr}$ are exchangeable, conditional on the emissions scenario.
Exchangeability implies that we hold the same prior beliefs about the output of every run from every model, given a particular scenario.
Therefore, we should specify the same probability model for each run of a particular scenario from every model.
We model the individual runs $X_{tmr}$ as
\begin{align}
  X_{Hmr} \mid X_{Hm} & \sim \Np{X_{Hm}}{\sigmahm^2}  &
  X_{Fmr} \mid X_{Fm} & \sim \Np{X_{Fm}}{\left( \sigmafm \sigmahm \right)^2} .
  \label{eqn:xtmr}
\end{align}
The outputs $X_{tmr}$ are assumed to be independent between runs $r$, conditional on the other parameters.
The model specific means $X_{tm}$ represent the expected climate of model $m$ at time $t$.
The model specific variances $\sigmahm^2$ quantify the spread of the runs from each model in the historical period, i.e., internal variability.
The coefficients $\sigmafm^2$ allow the internal variability of each model to change in the future period.
We assume that the historical and future periods are sufficiently separated in time that departures due to internal variability can be considered independent between periods.

In order to satisfy the assumption of exchangeability between the model outputs, we must also specify the same probability model for the expected climate of each model $X_{Hm}$ and $X_{Fm}$, and the internal variability of each model $\sigmahm^2$ and $\sigmafm^2$.
We model the expected climates as
\begin{align}
  X_{Hm}             & \sim \Np{\mu_H }{\sigmah^2}  &
  X_{Fm} \mid X_{Hm} & \sim \Np{\mu_F + \beta \left( X_{Hm} - \mu_H \right)}
                               {\sigma_{F \mid H}^2}
  \label{eqn:xsm}
\end{align}
and the internal variabilities as
\begin{align}
  \sigmahm^2 & \sim \Invgammap{\frac{\nu_H}{2}}
                              {\frac{\nu_H \esigmahm^2}{2}} &
  \sigmafm^2 & \sim \Invgammap{\frac{\nu_F}{2}}
                              {\frac{\nu_F \esigmafm^2}{2}}. 
  \label{eqn:taum}
\end{align}
The model specific parameters $X_{Hm}$, $X_{Fm}$, $\sigmahm^2$ and $\sigmafm^2$ are assumed to be independent between models $m$, conditional on the other parameters.
The common means $\mu_H$ and $\mu_F$ in \autoref{eqn:xsm} are interpreted as the representative climate of the ensemble in the historical and future periods respectively, i.e., representative in the sense that they summarize the climates simulated by the models.
The variances $\sigmah^2$ and $\sigmaf^2$ quantify the spread of the models around the representative climate, i.e., model uncertainty.
The parametrization of the internal variabilities in \autoref{eqn:taum} implies that $\esigmahm^2 = 1/\Ep{\sigmahm^{-2}}$, $\esigmafm^2 = 1/\Ep{\sigmafm^{-2}}$ and $\esigmafm^2 \esigmahm^2 = 1/\Ep{(\sigmafm \sigmahm)^{-2}}$.
Therefore, $\esigmahm^2$ and $\esigmafm^2$ can be interpreted as the representative internal variability of the ensemble.
The degrees-of-freedom $\nu_H$ and $\nu_F$ control the precision of $\sigmahm^2$ and $\sigmafm^2$ and quantify model uncertainty about the internal variability.

The parameter $\beta$ is intended to capture any linear association between the historical climates and future climate responses of the models, i.e., any emergent relationship, and is referred to as the \emph{emergent constraint}.
The emergent constraint applies to the expected climates of the models, not the individual runs, because emergent relationships are the result of model/process differences, not internal variability.
A value of $\beta = 1$ implies conditional independence of the expected response $X_{Fm} - X_{Hm}$ of model $m$ from its expected historical state $X_{Hm}$, i.e., $\Ep{X_{Fm} - X_{Hm} \mid X_{Hm}} = \mu_F - \mu_H$ for all $m$.
Any value of $\beta \neq 1$ implies that the expected historical climate $X_{Hm}$ is informative for the expected climate response $X_{Fm} - X_{Hm}$.

The representation in terms of the common unknown means, $\mu_H$ and $\mu_F$, induces a common prior correlation (dependence) between the model means and consequently the model outputs, e.g., $\covp{X_{Hm},X_{Hm^\prime}} = \varp{\mu_H}$ for all $m \neq m^\prime$.
Thus, we do not require the much stronger assumption that the model outputs are independent.

\subsection{The Earth system}
\label{sec:system}

Let $Y_{ta}$ represent the single actualization of the Earth system that we observe during time period~$t$.
We model the actualized climate as 
\begin{align}
  Y_{Ha}  \mid Y_H    & \sim \Np{Y_H}{\sigmaha^2}     &
  Y_{Fa}  \mid Y_F    & \sim \Np{Y_F}{\left( \sigmafa \sigmaha \right)^2}. 
  \label{eqn:yta}
\end{align}
The means $Y_H$ and $Y_F$ represent the expected climate of the Earth system in the historical and future periods respectively.
The variance $\sigmaha^2$ quantifies the historical natural variability in the Earth system, and the coefficient $\sigmafa^2$ represents any future change in variability.

Since each model attempts to approximate the Earth system as realistically as possible, we should hope that the both the expected climate and internal variability simulated by each model is informative climate of the Earth system.
While there are differences between individual climate models, both the mean climates and internal variabilities are usually similar to the Earth system, i.e., the observed quantities usually (but not always) lie within the range simulated by the different models \citep{Flato2013}.
We model the expected climate and natural variability of the Earth system conditional on the representative model as
\begin{align}
  Y_H                & \sim \Np{\mu_H}{\sigmadh^2}  &
  Y_F \mid Y_H       & \sim \Np{\mu_F + \beta \left( Y_H - \mu_H \right)}
                               {\sigmadf^2}
  \label{eqn:yt}
\end{align}
and
\begin{align}
  \sigmaha^2 & \sim \Invgammap{\frac{\nu_{Ha}}{2}}
                              {\frac{\nu_{Ha} \esigmahm^2}{2}} &
  \sigmafa^2 & \sim \Invgammap{\frac{\nu_{Fa}}{2}}
                              {\frac{\nu_{Fa} \esigmafm^2}{2}}.
  \label{eqn:taua}
\end{align}
In \autoref{eqn:yt}, we assume that the emergent constraint $\beta$ has a well understood physical basis, and therefore applies to the Earth system in the same way as the climate models.
The variances $\sigmadh^2$ and $\sigmadf^2$ quantify our uncertainty about the effects of common differences between the models and the Earth system, i.e., model inadequacy.
\autoref{eqn:taua} implies that $\Ep{1/\sigmaha^2} = 1/\esigmahm^2$ and $\Ep{1/\sigmafa^2} = 1/\esigmafm^2$.
The degrees-of-freedom $\nu_{Ha}$ and $\nu_{Fa}$ quantify model inadequacy in simulating natural variability in the Earth system.
In the language of \citet{Rougieretal2013}, the Earth system is assumed to be \emph{coexchangeable} with the models.
Conditioning on the representative model induces a correlation (dependence) between the expected climate and the model means, i.e., $\covp{Y_H,X_{Hm}} = \varp{\mu_H}$ for all $m$.

\subsection{The observed climate}
\label{sec:obs}

Let $Z_H$ be the observed climate during the historical period.
We model the observed climate as
\begin{equation}
  Z_H ~ \sim \Np{Y_{Ha}}{\sigmaz^2}.
  \label{eqn:z}
\end{equation}
The variance $\sigmaz^2$ quantifies our observation uncertainty.

\subsection{Making inferences about future climate}
\label{sec:inference}

The multi-model ensemble is described by nine parameters $\mu_H$, $\mu_F$, $\beta$, $\sigmah^2$, $\sigmaf^2$, $\esigmahm^2$, $\esigmafm^2$, $\nu_H$, and $\nu_F$.
Given outputs from a moderate number of climate models $M$, it should be possible to obtain reasonable inferences for the mean parameters $\mu_H$, $\mu_F$ and $\beta$, and the model uncertainty $\sigmah^2$ and $\sigmaf^2$.
The internal variability $\esigmahm^2$ and $\esigmafm^2$ can be distinguished from model uncertainty provided we have multiple initial condition runs from several models.
Some models may have only a single initial condition run in one or both time periods.
In that case, our hierarchical framework allows the model specific internal variability $\sigmahm^2$ and $\sigmafm^2$ to be estimated by borrowing strength from models with multiple runs, under the assumption that models should have similar internal variability (\autoref{eqn:taum}).
The most difficult parameters to infer are likely to be the degrees-of-freedom $\nu_H$ and $\nu_F$, since these are essentially variances of variances.

The Earth system is represented by a further four parameters $\sigmadh^2$, $\sigmadf^2$, $\nu_{Ha}$ and $\nu_{Fa}$.
The future parameters $\sigmadf^2$ and $\nu_{Fa}$ cannot be estimated from data, since we have no future observations of the Earth system.
Therefore, additional modeling assumptions are required.
If an estimate of the historical natural variability $\sigmaha^2$ is available, then this can be substituted directly, otherwise it can be inferred from the representative model using \autoref{eqn:taua}.

\subsubsection{Model inadequacy}

In principle, the historical model inadequacy quantified by $\sigmadh^2$ and $\nu_{Ha}$ could be estimated from a time series of observations and corresponding simulations.
This would require careful modeling to account for time-varying trends and to separate model inadequacy from internal variability and natural variability.
In addition, an \emph{extremely} long time series would be required, since the the discrepancy between the Earth system and the ensemble is expected to change only slowly over time.
Instead, we adopt the approach proposed by \citet{Rougieretal2013} and parameterize
the model inadequacy as proportional to the ensemble spread
\begin{equation}
  \begin{aligned}
      \sigmadh^2 & = \kappa^2 \sigmah^2  & \quad &&
      \sigmadf^2 & = \kappa^2 \sigmaf^2  \\
      \nu_{Ha}   & = \nu_H / \kappa^2  & \quad &&
      \nu_{Fa}   & = \nu_F / \kappa^2
  \end{aligned}
  \label{eqn:kappa}
\end{equation}
where $\kappa \geq 1$.
The coefficient $\kappa$ acts to inflate the ensemble spread in order to account for uncertainty due to processes not captured by any model, and errors common to all models.
Setting $\kappa = 1$ implies that the Earth system is exchangeable with the climate models, i.e., just another computer model.
The value of $\kappa$ must be fixed \textit{a priori}, and \citet{Rougieretal2013} suggest a value of $\kappa = 1.2$ for surface temperature.
Larger values of $\kappa$ might be appropriate for less well simulated processes, e.g., when the models are less informative for the real world and the observations lie outside of the spread in the models.

\subsubsection{Observation uncertainty}

Estimates of the observation uncertainty $\sigmaz^2$ are often not readily available.
Several modeling centers produce ``reanalysis'' products that combine multiple observation sources using complex data assimilation techniques and numerical weather models.
Given multiple reanalysis data sets we can approximate our uncertainty about the observed state of the climate.
Let $W_i$ be the output of reanalysis $i$, which we model as
\begin{equation}
  \label{eqn:w}
  W_i \sim \Np{\mu_W}{\sigma_W^2}
\end{equation}
where $\mu_W$ is interpreted as a representative reanalysis and the variance $\sigma_W^2$ quantifies the spread of the reanalyses.
We expect the representative reanalysis $\mu_W$ to be similar to the actualized climate $Y_{Ha}$, and so we model the representative reanalysis as
\begin{align}
  \label{eqn:yw}
  \mu_W & \sim \Np{Y_{Ha}}{\sigmadw^2}
\end{align}
The variance $\sigmadw^2$ quantifies our uncertainty about the discrepancy between the representative reanalysis and the actual climate, due to sparsity of observations, errors in the numerical weather models etc.
Similar to the models, we judge that the representative reanalysis is less like the actualized climate than the individual reanalyses are like the representative reanalysis, so we set
\begin{align}
  \sigmadw^2 & = \kappa_W^2 \sigma_W^2 & \kappa_W & \geq 1.
\end{align}
Conditioning the representative reanalysis $\mu_W$ on the actualized climate $Y_{Ha}$ in \autoref{eqn:yw} induces a correlation (dependence) between the models and the reanalyses, i.e, $\covp{W_i,X_{Hm}} = \varp{\mu_H} + \sigma_{\Delta_H}^2 + \sigma_a^2 + \sigma_{\Delta_W}^2$ for all $\lbrace i,m \rbrace$.
Such a correlation makes sense, since climate models and reanalyses are very closely related, sharing very similar numerical cores.

\section{Discussion and comparison to previous frameworks}
\label{sec:discussion}

The framework proposed in \autoref{sec:framework} was developed to model temperature data for which the normal distribution is a natural choice.
However, since the model outputs $X_{tmr}$ are assumed to be time-averages, e.g., 30-year means, the Central Limit Theorem guarantees that the distribution of the $X_{tmr}$ should converge to a normal distribution, regardless of the underlying distribution.
Therefore, the proposed framework should be suitable for a wide range of other climate variables. 
If necessary, different distributional choices can be substituted provided the hierarchical structure is respected in order to maintain the assumption of exchangeability between the model outputs.

In our application, the historical and future variables are the same, i.e., temperature. 
However, there are many examples of emergent relationships in the literature between different variables in the historical and future periods, e.g., \citet{Cox2018} relate historical temperature variability to Equilibrium Climate Sensitivity.
The framework proposed in \autoref{sec:framework} is easily generalized to the case of different historical and future variables by making the future internal variability independent of the historical internal variability in \autoref{eqn:xtmr}, and likewise the natural variability in \autoref{eqn:yta}, i.e. $\varp{X_{Fmr}} = \varphi_m^2$ rather than $\varp{X_{Fmr}} = \varphi_m^2 \sigma_m^2$.
No other changes are necessary since all other quantities are specified independently for historical and future variables. 

The formulation of the emergent relationship in Equations~\ref{eqn:xsm} and \ref{eqn:yt} reflects the linear relationships that have so far been documented in the literature.
\citet{Bracegirdle2012} also considered quadratic relationships and \citet{Halletal2019} propose the existence of more general functional relationships.
The methodology proposed here generalizes immediately to polynomial relationships and could easily be generalized to other parametric forms.

In Equations~\ref{eqn:xsm} and \ref{eqn:taum} we assume that the climate models are exchangeable, that is, they can be considered independent conditional on the representative model.
If we treat models that share common components as independent then we risk unfairly weighting particular groups of models.
Methods for assessing model dependence based on comparing spatial-temporal outputs have been shown to successfully capture similarities between groups of related models \citep{Masson2011,Knutti2013}.
However, current methods lack a formal statistical framework for combining projections from different models, and can produce unexpected results where models that are known to have little in common are considered close \citep{Sanderson2015a}.
\citet{Rougieretal2013} address the problem of model dependence by selecting a subset of models that they judge \textit{a priori} to be exchangeable.
We adopt a similar approach in \autoref{sec:results} based on readily available data about climate model structure and components shared between models.
By analyzing only a subset of the available data we risk losing valuable information.
However, the information loss is likely to be acceptable given the known similarities between many climate models \citep{Annan2011,Pennell2011}

The framework proposed here makes no assumptions about spatial dependence.
Climate model output is often analyzed grid box by grid box, and this is the approach we take in \autoref{sec:results}.
In practice, non-physical discontinuities between neighboring grid boxes are rarely a problem due to the inherent smoothness of computer model output in comparison to observations.
Accounting for spatial dependence could potentially lead to more efficient estimates by borrowing strength across neighboring grid boxes.
However, any increase in efficiency would come at the cost of additional complexity both in terms of the number of parameters and the computational requirements of fitting to all grid boxes simultaneously.
Several approaches have been proposed for modeling spatial structure in multi-model ensemble outputs, including harmonic basis functions \citep{Furrer2007}, kernel mixing \citep{Bhat2011}, and principal components \citep{Rougieretal2013,Sanderson2015a}.
However, it is not clear which (if any) of these approaches is most appropriate for multi-model experiments.
Differences in feature placement between models can result in overly smooth estimates that do not reflect the physical structure of the underlying field.
The methodology proposed here ensures that although posterior mean estimates may be over-smoothed in place, we retain uncertainty due to differences in feature placement thanks to explicit characterization of model uncertainty and inadequacy.

In the framework proposed here, we adopt the approach introduced by \citet{Chandler2013} and \citet{Rougieretal2013} and represent multi-model inadequacy as an unknown discrepancy between the climate system and a representative model.
This generalizes the well established single model approach in the uncertainty quantification literature \citep{Craig2001,Kennedy2001} by splitting the discrepancy into two parts: one common to all models, and one unique to each model.
The limitations of climate models in approximating the Earth system may manifest themselves in a variety of ways.
In the absence of stronger beliefs about how these limitations will manifest, an unknown discrepancy is the simplest and most intuitive way of representing the possibility.
Other approaches to representing model inadequacy in an ensemble of computer models may be possible, but we are not aware of any published alternatives.

In the introduction, we made the distinction between a purely statistical ``emergent relationship'', and an ``emergent constraint'' for which a plausible physical mechanism has been identified.
\citet{Halletal2019} make a similar distinction between what they call ``proposed'' and ``confirmed'' emergent constraints, and outline how a constraint might transition from ``proposed'' to ``confirmed''.
In formulating our framework, we assume that the emergent constraint applies in the Earth system the same way it does in the models.
This assumption is implicit in \emph{all} projections based on emergent constraints, although never stated.
By formulating a principled statistical framework, we make this assumption clear and transparent.
Thus, by making a projection based on an emergent relationship, we are making a strong statement of confidence in that relationship.
The framework proposed here addresses this by separating model inadequacy from model uncertainty, i.e., by allowing for additional uncertainty about the response of the Earth system.
However, the appropriate amount of additional uncertainty remains a subjective choice.

\subsection{Ensemble regression}

\citet{Bracegirdle2012} proposed a method for projection using emergent constraints known as ``ensemble regression''.
Ensemble regression is equivalent to simple linear regression of the model mean responses on the model mean historical climates, and can be written in our notation as
\begin{align*}
  \bar{X}_{Fm} - \bar{X}_{Hm} & \sim 
    \Np{\bar{X}_F - \bar{X}_H + \beta^\prime (\bar{X}_{Hm} - \bar{X}_H)}
       {\sigmaf^2}
\end{align*}
where $\bar{X}_{tm} = \sum_r X_{tmr} / R_{tm}$ and $\bar{X}_t = \sum_m \bar{X}_{tm} / M$.
This is equivalent to our \autoref{eqn:xsm} where $\beta^\prime = \beta - 1$, since $\Ep{\bar{X}_{tm}} = X_{tm}$ and $\Ep{\bar{X}_t} = \mu_t$.

Ensemble regression ignores uncertainty due to internal variability in the model means $\bar{X}_{Hm}$ and the ensemble mean $\bar{X}_H$.
It is well known that errors in the independent variable ($\bar{X}_{Hm} - \bar{X}_H$) in a regression will cause the slope estimate to be biased towards zero, a phenomenon known as \textit{regression dilution} or \textit{regression attenuation} \citep{Frost2000}.
Consider a balanced ensemble ($R_{Hm} = R_{Fm} = R \ \text{for all} \ m$) in which all models simulate the same internal variability in each time period, i.e.,
$\sigma_m^2 = \sigma^2 \ \text{and} \ \varphi_m^2 = 1 \ \text{for all} \ m$. 
The expected value of the linear regression estimate of the emergent constraint is
\begin{equation*}
  \Ep{\hat{\beta}^\prime} 
    = \frac{\covp{\bar{X}_{Fm} - \bar{X}_{Hm},\bar{X}_{Hm} - \bar{X}_H}}
           {\varp{                            \bar{X}_{Hm} - \bar{X}_H}}
    = \frac{\beta^\prime \sigmah^2 - \sigma^2 / R}
           {             \sigmah^2 + \sigma^2 / R}
\end{equation*}
where $\beta^\prime$ is the ``true'' value of the emergent constraint.
The bias is largest when the internal variability $\sigma^2$ is large compared to the model uncertainty $\sigmah^2$, or when the number of runs $R$ from each model is small.
The framework proposed in \autoref{sec:framework} avoids this bias by explicitly modeling internal variability and its relationship to the expected model climates $X_{tm}$.

In \citet{Bracegirdle2012}, the ensemble regression estimate of the response of the Earth system is
\begin{align*}
  Y_F - Y_H & \sim 
    \Np{\bar{X}_F - \bar{X}_H + \beta^\prime (Z_H - \bar{X}_H)}
       {\sigmaf^2}.
\end{align*}
This is equivalent to assuming the Earth system is exchangeable with the models and ignores the possibility of common differences between the models and the Earth system, as well as the effects of observation uncertainty and natural variability.
The framework proposed here explicitly allows for common model inadequacy, observation uncertainty and natural variability.

\subsection{A simple hierarchical framework}

\citet{Bowman2018} propose a hierarchical framework for emergent constraints without explicit reference to climate models. In our notation, the linear normal-theory version is
\begin{align*}
  Y_H          & \sim \Np{\mu_H}{\sigmah^2} &
  Y_F \mid Y_H & \sim \Np{\mu_F + \beta (Y_H - \mu_H)}{\sigmaf^2}
\end{align*}
and $Z_H \mid Y_H \sim \Np{Y_H}{\sigmaz^2}$. In practice, the parameters $\mu_H$, $\mu_F$, $\beta$, $\sigmah$ and $\sigmaf$ are estimated from an ensemble of climate models by assuming
\begin{align*}
  X_{Hm}             & \sim \Np{\mu_H}{\sigmah^2} &
  X_{Fm} \mid X_{Hm} & \sim \Np{\mu_F + \beta (X_{Hm} - \mu_H)}{\sigmaf^2}
\end{align*}
for all $m = 1,\ldots,M$. This is identical to our \autoref{eqn:xsm}, so the framework proposed by \citet{Bowman2018} is almost equivalent to Ensemble Regression \citep{Bracegirdle2012}, but allowing for observation uncertainty. However, the inclusion of the prior on $Y_H$ implies that the posterior expectation of $Y_F$ \citep[Eqns~11,13,17]{Bowman2018} is
\begin{align*}
  \Ep{Y_F \mid Z_H} = \mu_F + \beta \frac{\sigmaz^{-2}}{\sigmaz^{-2} + \sigmah^{-2}} \left( Z_H - \mu_H \right).
\end{align*}
So the expected future climate $Y_F$ experiences a shrinkage towards the representative climate $\mu_F$ depending on how informative the observations are compared to the models for the historical climate $Y_H$, i.e., the ratio of $\sigmaz^2$ to $\sigmah^2$.
No attempt is made to account for model inadequacy, the Earth system is implicitly assumed to be exchangeable with the models. Further, only one run from each model is used, thus ignoring internal variability and leaving the estimated emergent relationship vulnerable to regression dilution.

\subsection{The coexchangeable framework}

\citet{Rougieretal2013} propose a model of the joint distribution of the historical and future climate in multi-model experiments known as the coexchangeable framework.
In our notation
\begin{align*}
  \bX_m & \sim \Np{\bmu}{\bSigma       } & m & = 1,\ldots,M \\
  \bY   & \sim \Np{\bA \bmu}{\bSigma_\Delta} &
    Z_H & \sim \Np{Y_H }{\sigmaz^2   }
\end{align*}
where $\bX_m = (X_{Hm},X_{Fm})^T$, $\bY = (Y_H,Y_F)^T$, $\bmu = (\mu_H,\mu_F)^T$.
The matrix $\bA$ is assumed known and allows for transformation of variables between model world and the real world (the default choice is $\bA = \bI$, the identity).
The exchangeable framework is a special case of the the coexchangeable framework where $\bSigma_\Delta = \bSigma$ and $\bA = \bI$.
The framework proposed here is an extension of the coexchangeable framework with $\bA = \bI$ and
\begin{align*}
  \bmu & =
    \begin{pmatrix}
      \mu_H  \\ 
      \mu_F 
    \end{pmatrix} & 
  \bSigma & =
    \begin{pmatrix}
            \sigmah^2 & \beta \sigmah^2 \\ 
      \beta \sigmah^2 & \beta^2 \sigmah^2 + \sigmaf^2
    \end{pmatrix}.
\end{align*}
However, the basic coexchangeable framework does not distinguish between model differences and internal variability, and does not account for natural variability in the Earth system.
The extended framework proposed here accounts for both of these additional sources of uncertainty.

\citet{Rougieretal2013} suggest the following parametrization of the model inadequacy
\begin{align*}
  \bSigma_\Delta & = \kappa^2 \bSigma + \bD
\end{align*}
where $\bD$ is a diagonal matrix with $diag(\bD) = (D_H^2,D_F^2)^T$.
The variances $D_H^2$ and $D_F^2$ are intended to guard against overly precise projections when models are in close agreement.
However, this parametrization has unexpected consequences for emergent constraints.
Standard results for the multivariate normal distribution show that
\begin{equation*}
  \Ep{Y_F \mid Y_H} = \mu_F + \beta^\star ( Y_H - \mu_H )
  \quad \text{where} \quad
  \beta^\star = \frac{\covp{Y_F,Y_H}}{\varp{Y_H}} 
              = \frac{\kappa^2 \sigmah^2}{\kappa^2 \sigmah^2 + D_H^2} \beta
\end{equation*}
The emergent constraint shrinks towards zero by an amount that depends on $D_H^2$.
This is difficult to defend given that we have assumed the emergent constraint has a physical basis and should apply to the Earth system.
Similar terms $D_H^2$ and $D_{F \mid H}^2$ could be added to our \autoref{eqn:kappa}, but without effecting the emergent constraint, since then $\covp{Y_F,Y_H} = \varp{Y_H} = \kappa^2 \sigmah^2 + D_H^2$ and $\beta^\star = \beta$.
The difference is due to our formulation in terms of conditional rather than marginal variances.
Like $\sigmadh^2$ and $\sigmadf^2$, $D_H^2$ and $D_{F \mid H}^2$ are difficult to specify \textit{a priori} without additional data. 
One possibility might be to consider the spread of a family of closely related models as a lower bound for the model inadequacy.

\subsection{The generalized truth-plus-error framework}

\citet{Chandler2013} proposed an alternative joint framework for multi-model projection
\begin{align*}
  \bX_{mr} & \sim \Np{\bX_m}{\bSigma_m     } & r & = 1,\ldots,N_m \\ 
  \bX_m    & \sim \Np{\bmu }{\bLambda_m    } & m & = 1,\ldots,M   \\
  \bmu     & \sim \Np{\bY  }{\bSigma_\Delta} &
    Y_{Ha} & \sim \Np{Y_H  }{\sigmaha^2    }
\end{align*}
where $\bX_{mr} = (X_{Hmr},X_{Fmr})^T$, $\bX_m = (X_{Hm},X_{Fm})^T$, $\bmu = (\mu_H,\mu_F)^T$, $\bY = (Y_H,Y_F)^T$. 
The variances $\bLambda_{tm}$ represent the propensity of each simulator to deviate from the ensemble consensus.
This provides flexibility to incorporate prior knowledge that certain climate models are more or less similar to each other.
Internal variability and model inadequacy are both accounted for.
In contrast to \citet{Rougieretal2013}, natural variability is accounted for, but observation uncertainty is ignored.

\citet{Chandler2013} suggests estimating the historical model inadequacy from data as $\sigmadh^2 = (Y_{Ha} - \mu_H)^2$ then setting
\begin{align*}
  \bSigma_\Delta & =
    \begin{pmatrix}
      \sigmadh^2 &              \sigmadh^2 \\ 
      \sigmadh^2 & (1 + \kappa) \sigmadh^2
    \end{pmatrix}
\end{align*}
for $\kappa > 0$.
This parametrization ignores any emergent constraints in the projection of the future climate.
In addition, estimating $\sigmadh^2$ from a single observation $Y_{Ha}$ provides very limited information and makes the analysis vulnerable to outlying or spurious measurements.

The frameworks proposed by \citet{Chandler2013} and \citet{Rougieretal2013} are conceptually very different and appear incompatible.
The most obvious difference is the direction of conditioning between the system $\bY$ and the representative or consensus climate $\bmu$.
However, \citet{Rougieretal2013} demonstrated that a simplified form of the generalized truth-plus-error framework can be viewed as a special case of the coexchangeable framework (up to the second moments), for particular choices of $\bA \neq \bI$ and $\bSigma_\Delta$.
It is interesting to note that when all the distributions are normal, identical priors are set for related quantities and $\bA = \bI$, both frameworks produce identical posterior inferences.
This is not the case when the assumption of normality is relaxed, and should not be interpreted as meaning that both formulations are equivalent and can be used interchangeably.

\citet{BerlinerKim2008} also considered the direction of conditioning between climate models the Earth system and concluded that it should be decided by our ability to formulate the relevant distributions, to interpret them, and to perform the necessary computations.
We find it more natural to consider the actual climate as the sum of our knowledge (the representative model) plus what we do \emph{not} understand (model inadequacy), than vice-versa.
Hence we adopt a coexchangeable representation for the models.
In contrast, we use a truth-plus-error representation for the reanalyses in \autoref{eqn:yw}.
This feels more natural since the reanalyses are trying to approximate an observable ($Y_{Ha}$), rather than an abstract quantity (``the climate'') for the models.

\subsection{Reliability ensemble averaging}

\citet{Tebaldi2005} proposed a probabilistic interpretation of the heuristic ``reliability ensemble averaging'' framework of \citet{Giorgi2003}.
The framework belongs to the truth-plus-error family, with some interesting features.
Multivariate extensions were proposed by \citet{Smith2009} and \citet{Tebaldi2009}, and a similar spatial framework was proposed by \citet{Furrer2007}.
The basic framework in our notation is given by
\begin{align*}
  X_{Hm}             & \sim \Np{Y_H}{\lambda_m^2} &
  X_{Fm} \mid X_{Hm} & \sim \Np{Y_F + \beta (X_{Hm} - Y_H)}
                               {(\theta \lambda_m)^2} \\
  Y_{Ha}             & \sim \Np{Y_H}{\sigmaha^2}.
\end{align*}
Similar to \citet{Chandler2013}, the model climates $X_{tm}$ are conditioned on the Earth system climate $Y_t$, and the variances $\lambda_m$ are interpreted as the propensity of each model to deviate from the system.
The coefficient $\theta$ allows the propensity of the models to differ from the system to change in the future period.
Somewhat confusingly, natural variability in the Earth system is accounted for, but internal variability in the models is ignored.
Observation uncertainty and model inadequacy are also both neglected.

The framework proposed by \citet{Tebaldi2005} includes something similar to an emergent constraint.
It is instructive to consider this alternative formulation in detail.
The expectation of the full conditional posterior distribution of future climate \citep[Eqn.~A9]{Tebaldi2005} is
\begin{align*}
  \Ep{Y_F \mid \ldots} 
    & = \frac{\sum_m \lambda_m^{-2} X_{Fm}}{\sum_m \lambda_m^{-2}}
      + \beta \left( Y_H - \frac{\sum_m \lambda_m^{-2} X_{Hm}}{\sum_m \lambda_m^{-2}} \right).          
\end{align*}
This is equivalent to our \autoref{eqn:yt}, if $\lambda_m^2 = \sigmah^2 \ \text{for all} \ m$, i.e., if all the models are exchangeable.
Let $\lambda_m^2 = \sigmah^2$ and $\theta \lambda_m^2 = \sigmaf^2$ for all $m$, then the posterior expectation of $Y_H$ \citep[Eqn.~A8]{Tebaldi2005} is
\begin{align*}
  \Ep{Y_H \mid \ldots} 
    & = \frac{\sigmaha^{-2} Y_{Ha} + M \left( \sigmah^{-2} \bar{X}_H +  \sigmaf^{-2} \beta ( Y_F - \bar{X}_F + \beta \bar{X}_H ) \right)}
             {\sigmaha^{-2} + M \left( \sigmah^{-2} + M \sigmaf^{-2} \beta^2 \right)}
\end{align*}
In comparison, the posterior expectation of $Y_H$ in our framework is
\begin{align*}
  \Ep{Y_H \mid \ldots}  & =
    \frac{\sigmaha^{-2} Y_{Ha} + \sigmah^{-2} \mu_H + 
                \sigmaf^{-2} \beta \left( Y_F - \mu_F + \beta \mu_H \right)}
             {\sigmaha^{-2} + \sigmah^{-2} + \sigmaf^{-2} \beta^2}
\end{align*}
assuming $\kappa = 1$, i.e., the models are exchangeable with the Earth system.
Both estimates are weighted averages of the model outputs and the actualized climate $Y_{Ha}$.
The two estimates effectively differ only in the weight given to the models.
Under the framework proposed by \citet{Tebaldi2005}, the models receive $M$ times more weight than under our framework.
As a result, the posterior expectation of the expected climate $Y_H$, and consequently the projected climate $Y_F$, will lie much closer to the consensus climate, and approach the consensus as the number of models increases.
In fact, the framework proposed by \citet{Tebaldi2005} implies that we can learn the expected climate $Y_H$ and $Y_F$ to any degree of precision we require, simply by adding more climate models \citep{Lopez2006}.
Given the existence of shared errors in all climate models, such an assumption is unsupportable.
\citet{Tebaldi2009} later proposed the inclusion of a common model bias term to address this issue.
However, the common bias was treated as a fixed quantity to be estimated, and does not contribute to our uncertainty about the Earth system in the same way as the model inadequacy terms proposed here and by \citet{Rougieretal2013} and \citet{Chandler2013}.

\subsection{Model weighting and Bayesian model averaging}

A variety of model weighting schemes have been proposed in the literature (see \autoref{sec:intro} for examples), but all have essentially the same functional form
\begin{align*}
  Y_t = \sum_{m=1}^M w_m X_{tm}.
\end{align*}
The actual climate (or climate response) $Y_t$ is modeled as a weighted combination of the model outputs.
Depending on the exact formulation, the weights $w_m$ may be constrained to be positive and sum to one.
The weights $w_m$ are estimated by comparing observations of the historical climate $Y_H$ with model simulations $X_{Hm}$ of the same period.
The same weights are then applied to future simulations $X_{Fm}$ to obtain projections of the future climate $Y_F$ or climate response $Y_F - Y_H$.
Bayesian Model Averaging differs from simple model weighting by dressing each simulation $X_{tm}$ with a kernel, so $Y_t$ becomes a mixture model.

In principle, model weighting will respect emergent relationships.
Consider the example of \autoref{fig:example}.
If the models closest to the observations receive the most weight, then the projected climate response will be lower than the ensemble mean estimate.
However, unless the models further from the observations receive almost zero weight, the projected response will shrink towards the ensemble mean.
The amount of shrinkage will depend on the exact form of the weights.
In practice, the weights $w_m$ are usually estimated by comparing model performance at multiple locations, often across the entire study region \citep[e.g.,][]{Bhat2011,Knutti2017}.
If the emergent relationship does not apply across the entire region, or varies within the region, then the weights are unlikely to reflect the relationship and the constraining behavior will be lost.

\section{Application to Arctic climate change}
\label{sec:results}

The CMIP5 ensemble includes output from more than 40 models submitted by over 20 centers around the world.
In order to satisfy the assumption of exchangeability in \autoref{sec:framework}, we consider a subset of the models that we judge to be approximately exchangeable.
The 13 chosen models and the number of runs available from the historical and future periods are listed in \autoref{tab:models}.
The models included in the thinned ensemble were chosen to have similar horizontal and vertical resolutions, but to minimize common component models according to the detailed information in Table 9.A.1 of \citet{Flato2013}.
In particular, only one model was retained from any one modeling center, generally the most recent and feature complete version submitted by each center.
Full details of the thinning process are given in the supplementary material.
Our approach to ensemble thinning differs from that of \citet{Rougieretal2013} who chose models judged to be most similar to a familiar model, effectively minimizing the differences between the models.
In contrast, by choosing models with the fewest common components, we are effectively maximizing the differences between the models.
In doing so, we aim to capture the broadest range of uncertainty due to model differences.

\subsection{Prior modeling and posterior computation}

\begin{table}[t!]
  \caption{Prior probability distributions.}
  \label{tab:priors}
  \centering
    \begin{small}
      \begin{tabular}{lcc}
        \hline \hline
        Description & Parameters & Prior \\
        \hline
        Representative historical climate   & 
          $\mu_H$                           & $\Np{0}{10^6}$ \\
        Representative future climate       & 
          $\mu_F$                           & $\Np{\mu_H}{10^6}$ \\
        Emergent constraint                 & 
          $\beta$                           & $\Np{1}{10^6}$ \\
        Representative Internal variability & 
          $\psi^2, \theta^2$                & $\Invgammap{10^{-3}}{10^{-3}}$ \\
        Model uncertainty                   & 
          $\sigma_H^2, \sigma_{F \mid H}^2$ & $\Invgammap{10^{-3}}{10^{-3}}$ \\
        Degrees-of-freedom                  & 
          $\nu_H, \nu_F$                    & $\Expp{1/M}$ \\
        Reanalysis uncertainty              & 
          $\sigma_W^2$                      & $\Invgammap{10^{-3}}{10^{-3}}$ \\
        \hline
      \end{tabular}
    \end{small}
\end{table}

Before computing posterior projections of late 21st century warming in the Arctic, we need to specify prior distributions for all unknown parameters.
For consistency, we adopt the assessment made by \citet{Rougieretal2013} and set $\kappa = \kappa_W = 1.2$.
Vague conjugate prior probability distributions were specified for the parameters and are listed in \autoref{tab:priors}.
The resulting full conditional posterior distributions all have standard forms with the exception of the degrees-of-freedom $\nu_H$ and $\nu_F$.
Therefore, posterior inference can be efficiently accomplished by Gibbs' sampling with Metropolis-Hastings steps for $\nu_H$ and $\nu_F$.
Full details of the posterior sampling procedure are given in the Supplementary Material.

Posterior analysis was performed for each grid box separately.
Identical priors were specified at all grid boxes.
Four parallel chains were initialized for each grid box, from over-dispersed starting points.
Initially, \num{20000} samples were performed by each chain for each grid box.
The first \num{10000} samples were discarded as burn-in, and Gelman-Rubin diagnostics performed on the remaining \num{10000} samples \citep{Gelman1992}.
If any random quantity had a potential scale reduction factor greater than 1.10, then sampling was continued for a further \num{10000} samples per chain and diagnostics performed again until satisfactory convergence was indicated.
We store every 40th sample from the last \num{10000} samples of each chain, leading to a final sample size of \num{1000} for each grid box.
The Metropolis-within-Gibbs' sampler was implemented in the R statistical computing language \citep{R}.
Computation time for four parallel chains of \num{20000} samples at a single grid box is around \SI{5.5}{\second} on a standard Linux workstation.
The samplers for all grid boxes converged successfully.
Convergence was achieved after the initial \num{20000} samples at \SI{50}{\percent} of grid boxes.
Less than \SI{2}{\percent} of grid boxes required more than \num{100000} samples before convergence.

\subsection{Model checking}

Inspection of the posterior distributions showed that, despite the small ensemble size, the ensemble parameters $\mu_H$, $\mu_F$, $\beta$, $\sigmah^2$, $\sigmaf^2$ and $\esigmahm^2$ and $\esigmafm^2$ are all very well constrained by the data.
As expected, the degrees-of-freedom $\nu_H$ and $\nu_F$ were only mildly constrained compared to the exponential prior. 
The inter-quartile range (IQR) for the mode of $\nu_H$ over the \num{2880} grid boxes was \numrange{5}{10}, and for $\nu_F$ the IQR was \numrange{5}{8}, compared to the mode of zero for the exponential prior.
However, both $\nu_H$ and $\nu_F$ tended to have long tails at individual grid boxes.
Due to the extremely small sample size, the reanalysis spread $\sigma_W$ was relatively poorly constrained compared to the other parameters, but the posterior mean was below \SI{2.0}{\celsius} at more than \SI{75}{\percent} of grid boxes.

Monte Carlo standard errors were computed for each parameter at each grid box \citep{Flegal2008,mcmcse}.
The Monte Carlo standard error rarely accounted for more than \SI{4.3}{\percent} of the posterior standard error, or exceeded \SI{3.8}{\percent} of the absolute posterior mean.

Examination of correlation matrices for the posterior samples revealed that only  the means $\mu_H$ and $\mu_F$ are consistently highly correlated (IQR $\corrp{\mu_H,\mu_F}$ \numrange{0.69}{0.93}), which is to be expected given the relationship in \autoref{eqn:xsm}.
Unsurprisingly, the internal variability $\esigmahm^2$ and $\esigmafm^2$ are also moderately correlated (IQR \numrange{-0.44}{-0.37}).
The only other parameters to have consistently non-zero correlation in the posterior samples were $\esigmahm^2$ and $\nu_H$ (IQR \numrange{0.13}{0.28}), and $\esigmafm^2$ and $\nu_F$ (IQR \numrange{0.08}{0.16}).
Again, this is not surprising given the close relationship between these parameters in \autoref{eqn:taum}, and the small number of initial condition runs available from each model.
None of these findings is particularly troubling, and so we conclude that the posterior simulation worked well.

\begin{figure}[t]
  \centering
  \begin{subfigure}{0.45\textwidth}
    \subcaption{} \label{fig:cv-hist}
    \includegraphics[width=\textwidth]{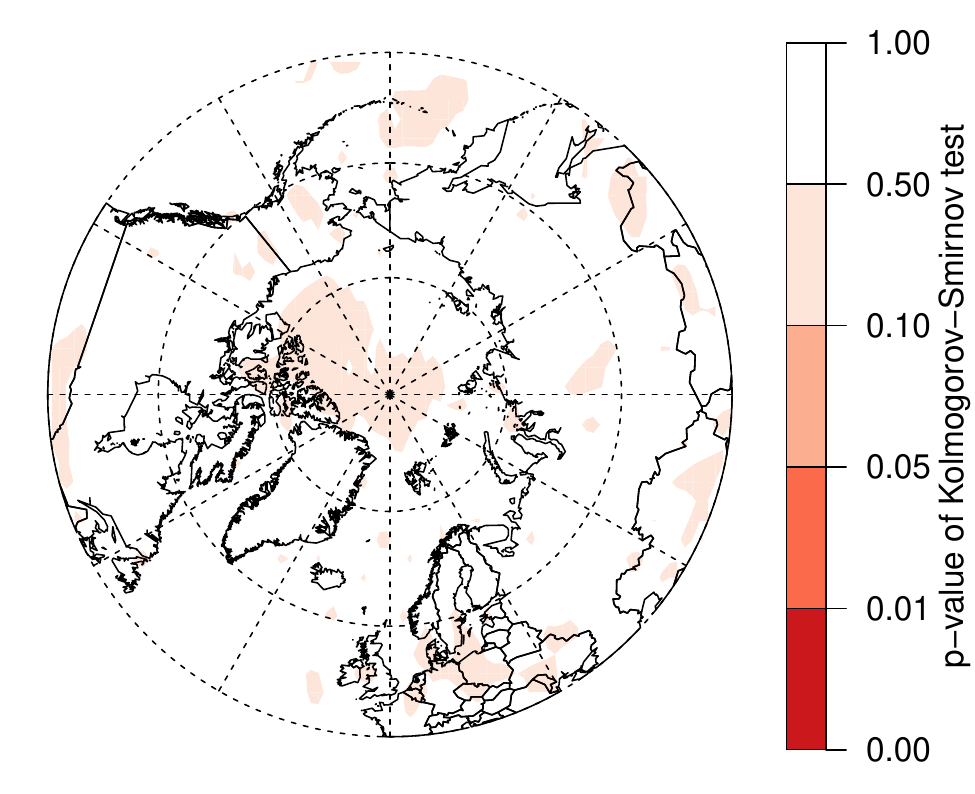}
  \end{subfigure}
  \begin{subfigure}{0.45\textwidth}
    \subcaption{} \label{fig:cv-fut}
    \includegraphics[width=\textwidth]{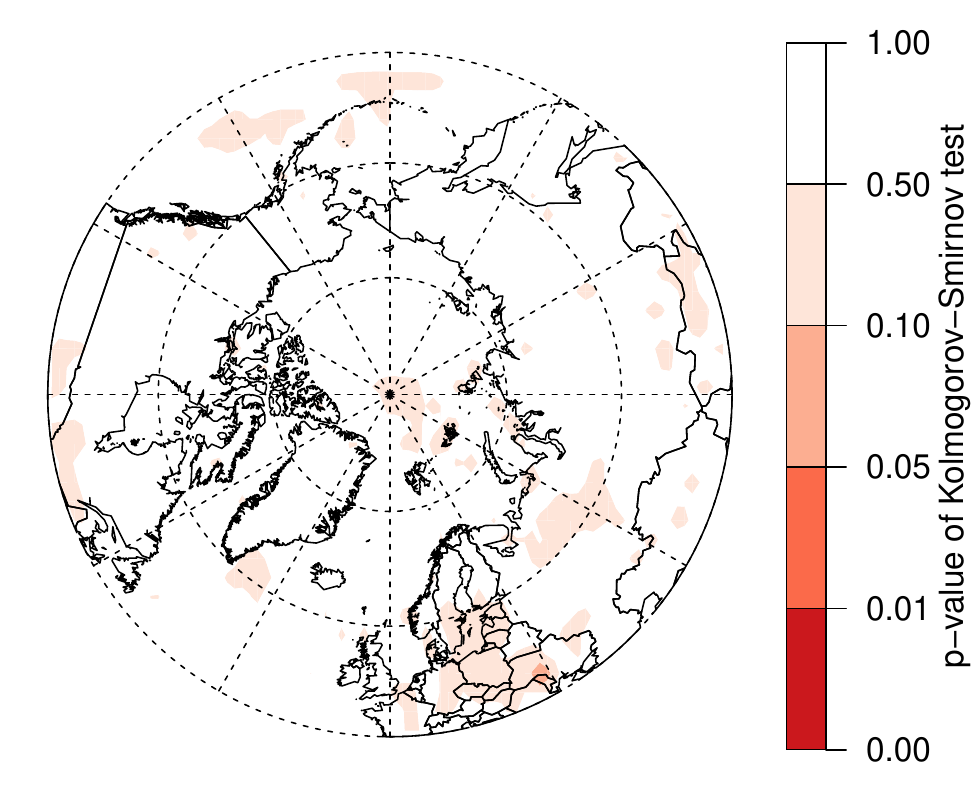}
  \end{subfigure}
   \caption{
     Cross validation.
     p-values of Kolmogorov-Smirnov tests for uniformity of leave-one-out cross-validated climate response predictions leaving out
     (\subref*{fig:cv-hist}) 
      all data; and
     (\subref*{fig:cv-fut}) 
       only future data from each model in turn.
  }
  \label{fig:cv}
\end{figure}

We checked the assumption of exchangeability between models using a leave-one-out cross-validation approach similar to \citet{Smith2009} and \citet{Rougieretal2013}.
Each model in turn is left out of the analysis, and the expected response $X_{Fm}^\star - X_{Hm}^\star$ of a new model is predicted.
The predictions are compared to the model output using a probability integral transform, i.e., by computing the probability that the response under the leave-one-out predicted distribution is less than the mean response of the excluded model.
If the models are exchangeable, then the distribution over the models of the transformed projections should be uniform.
Kolmogorov-Smirnov tests were used to assess uniformity at each grid box (see \autoref{fig:cv}).
A small amount of non-uniformity is expected due to shrinkage of the representative climate towards the observations.
In \autoref{fig:cv-hist}, we withhold all data from each model in turn.
There is no evidence against the null hypothesis that the models are exchangeable.
No grid boxes are significantly non-uniform at the \SI{10}{\percent} level.
In \autoref{fig:cv-fut}, we withhold only the future simulations to test  conditional exchangeability given any emergent relationships.
Only two grid boxes are significantly non-uniform at the \SI{10}{\percent} level.
The cross-validation procedure suggests that the chosen models can be considered exchangeable.

\subsection{Results}

\begin{figure}[t]
  \begin{subfigure}{0.32\textwidth}
    \subcaption{} \label{fig:yh}
    \includegraphics[width=\textwidth]{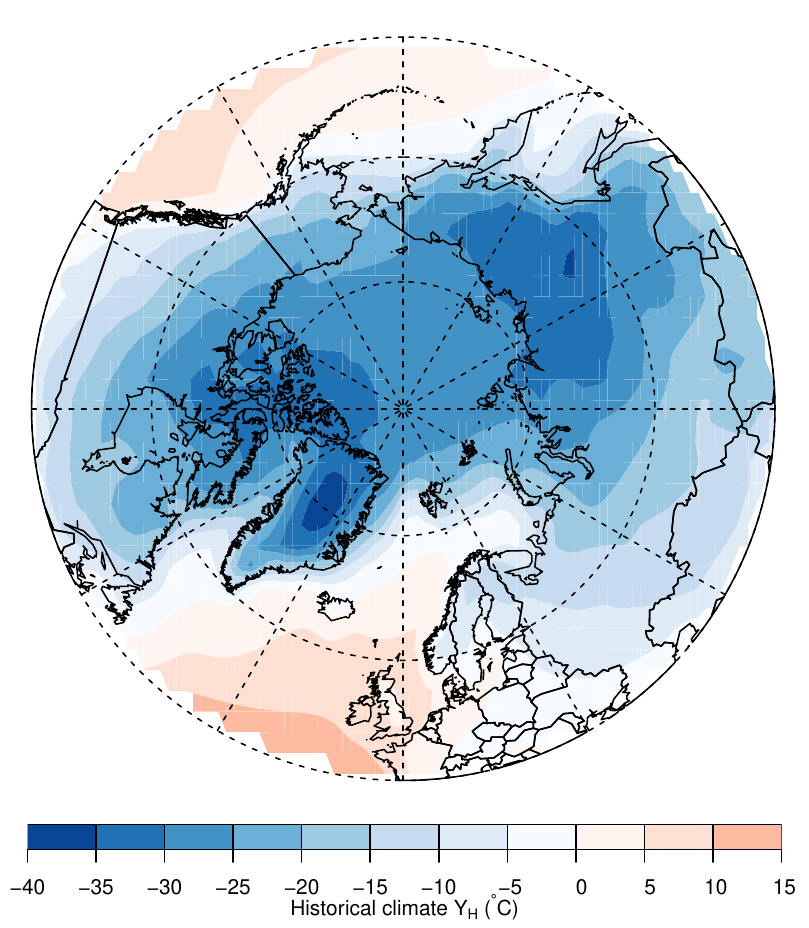}
  \end{subfigure}
  \begin{subfigure}{0.32\textwidth}
    \subcaption{} \label{fig:yf}
    \includegraphics[width=\textwidth]{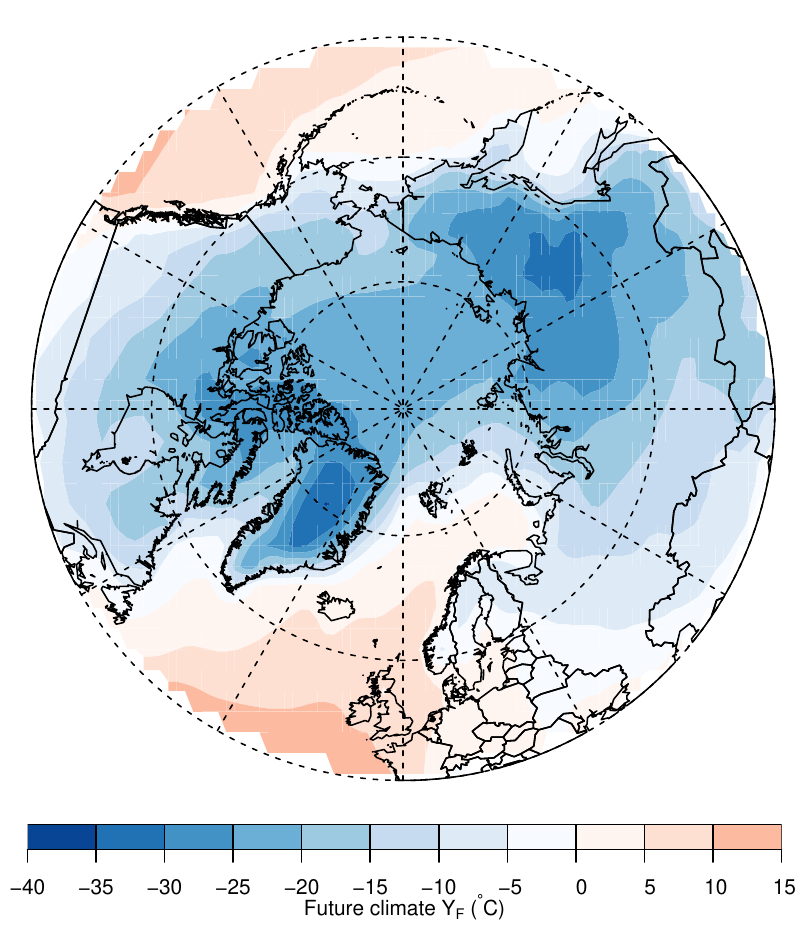}
  \end{subfigure}
  \begin{subfigure}{0.32\textwidth}
    \subcaption{} \label{fig:yr}
    \includegraphics[width=\textwidth]{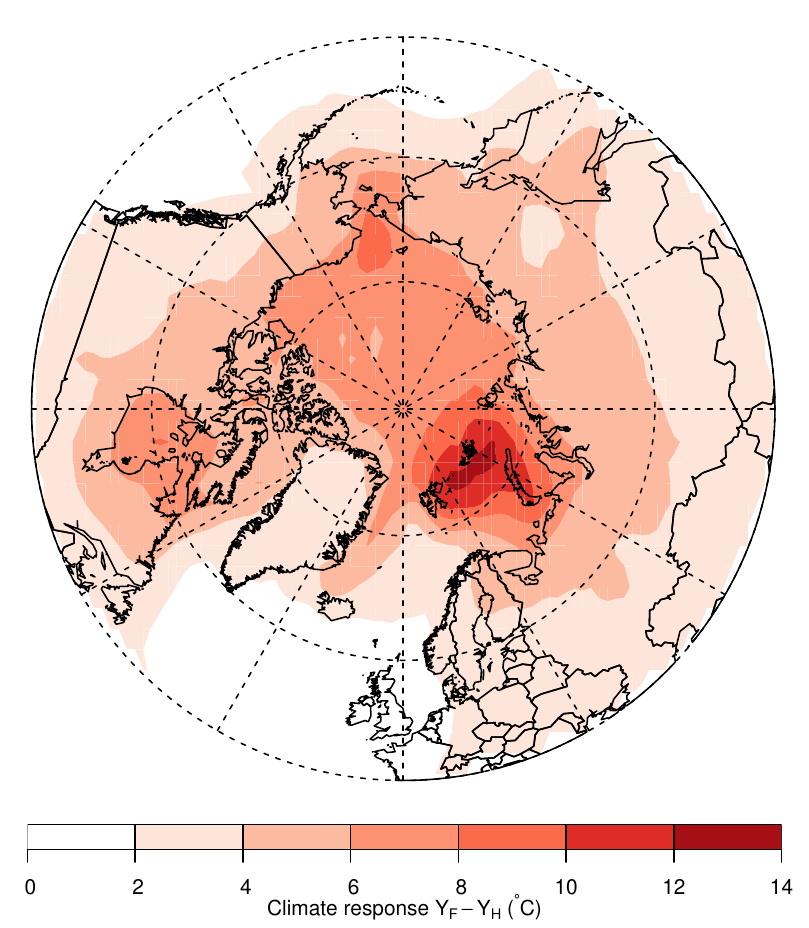}
  \end{subfigure}
  \caption{
    Expected climate.
    The posterior mean of
    (\subref*{fig:yh}) 
      the historical climate $Y_H$;
    (\subref*{fig:yf}) 
      the future climate $Y_F$; and
    (\subref*{fig:yr}) 
      the climate response $Y_F - Y_H$.}
  \label{fig:y}
\end{figure}

The posterior mean estimates of the expected historical climate $Y_H$, future climate $Y_F$, and climate response $Y_F - Y_H$ are shown in \autoref{fig:y}.
The \SI{0}{\celsius} contour that approximates the sea ice edge has receded noticeably in the projected future climate $Y_F$ in \autoref{fig:yf} compared to the historical climate $Y_H$ in \autoref{fig:yh}.
The projected warming tends to increase with latitude in \autoref{fig:yr}.

\begin{figure}[!ht]
  \centering
  \begin{subfigure}{0.45\textwidth}
    \subcaption{} \label{fig:deltah}
    \includegraphics[width=\textwidth]{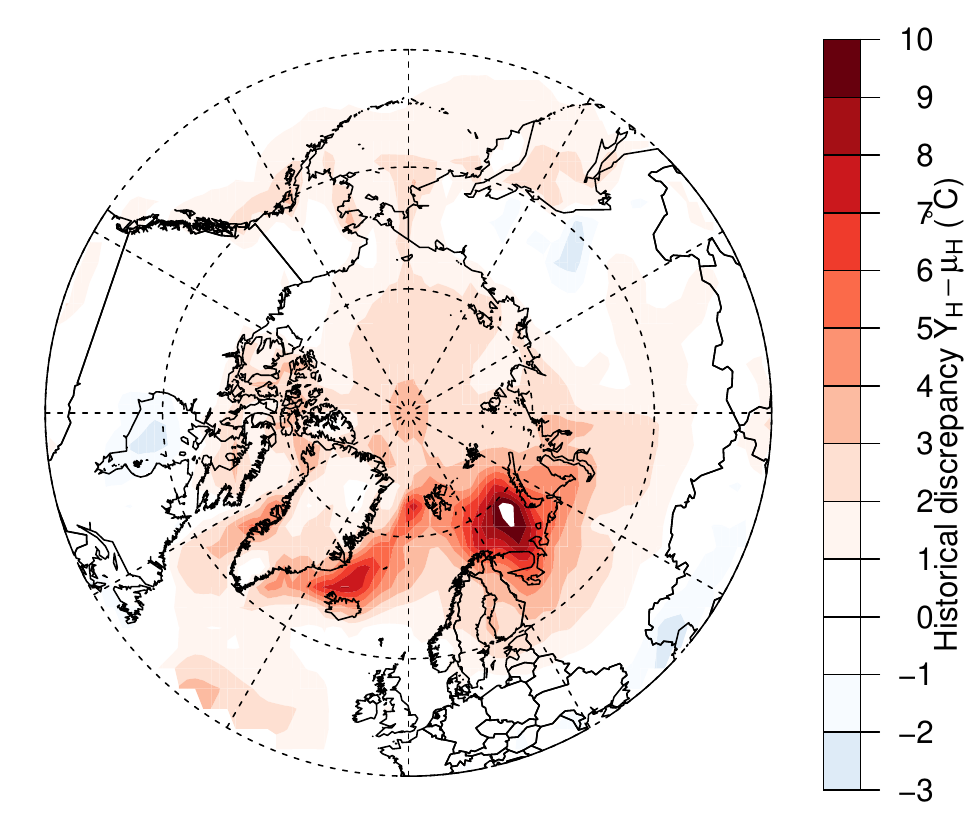}
  \end{subfigure}
  \begin{subfigure}{0.45\textwidth}
    \subcaption{} \label{fig:beta}   
    \includegraphics[width=\textwidth]{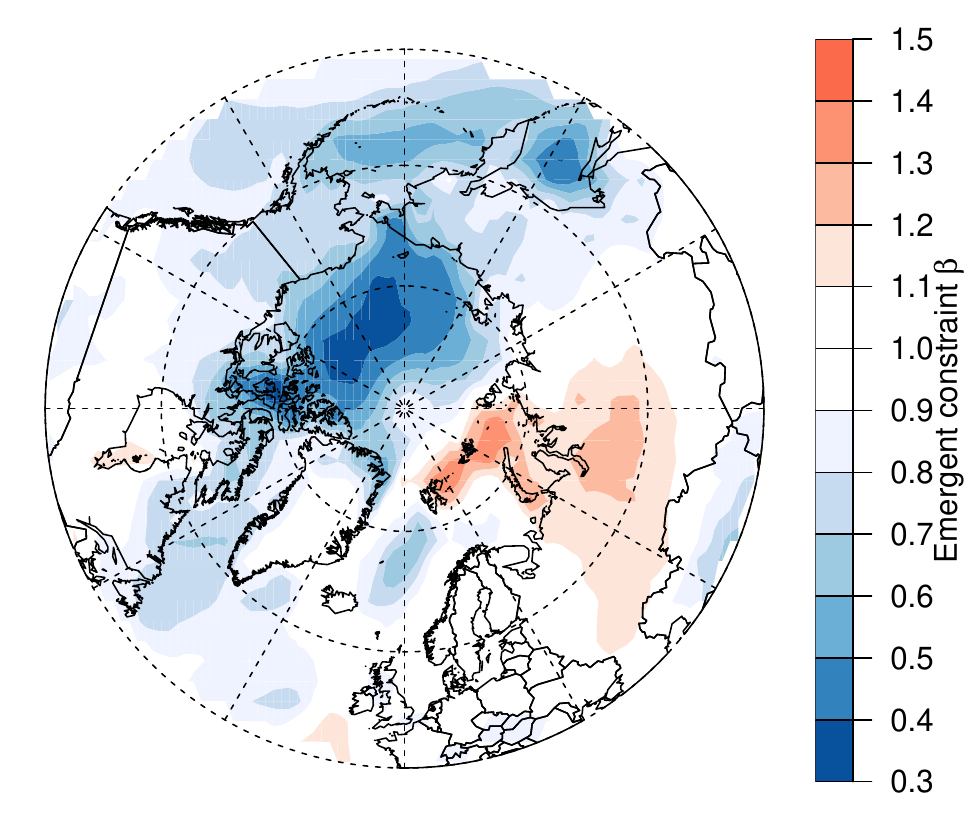}
  \end{subfigure} \\
  \begin{subfigure}{0.45\textwidth}
    \subcaption{} \label{fig:diff}
    \includegraphics[width=\textwidth]{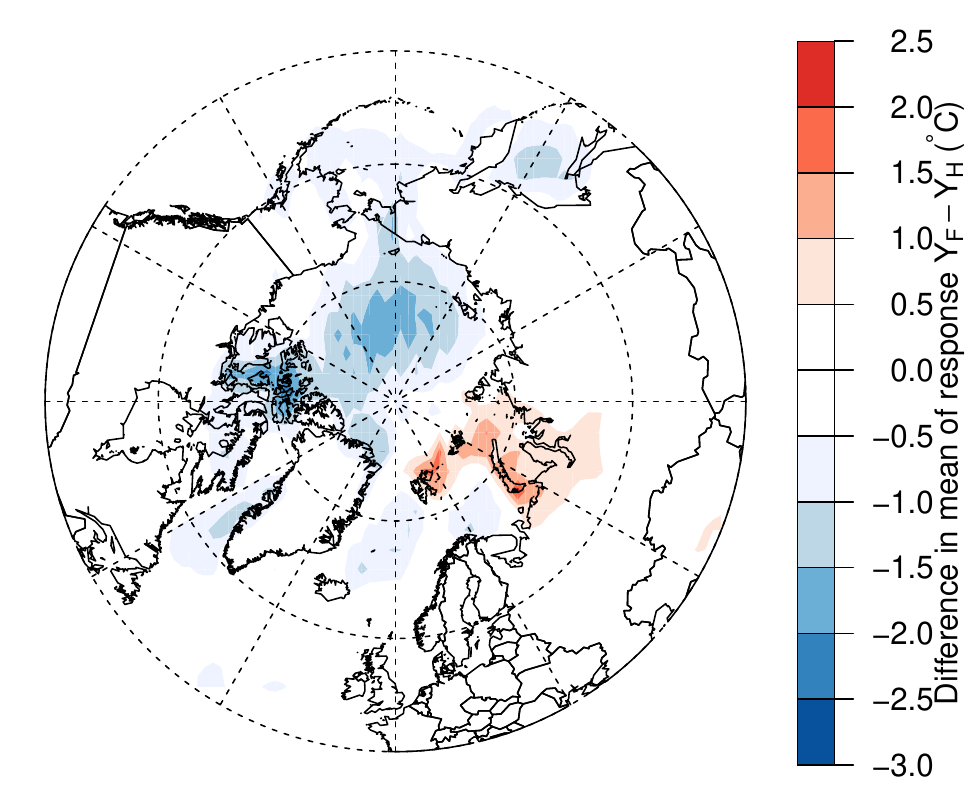}
  \end{subfigure}
  \begin{subfigure}{0.45\textwidth}
    \subcaption{} \label{fig:ratio}
    \includegraphics[width=\textwidth]{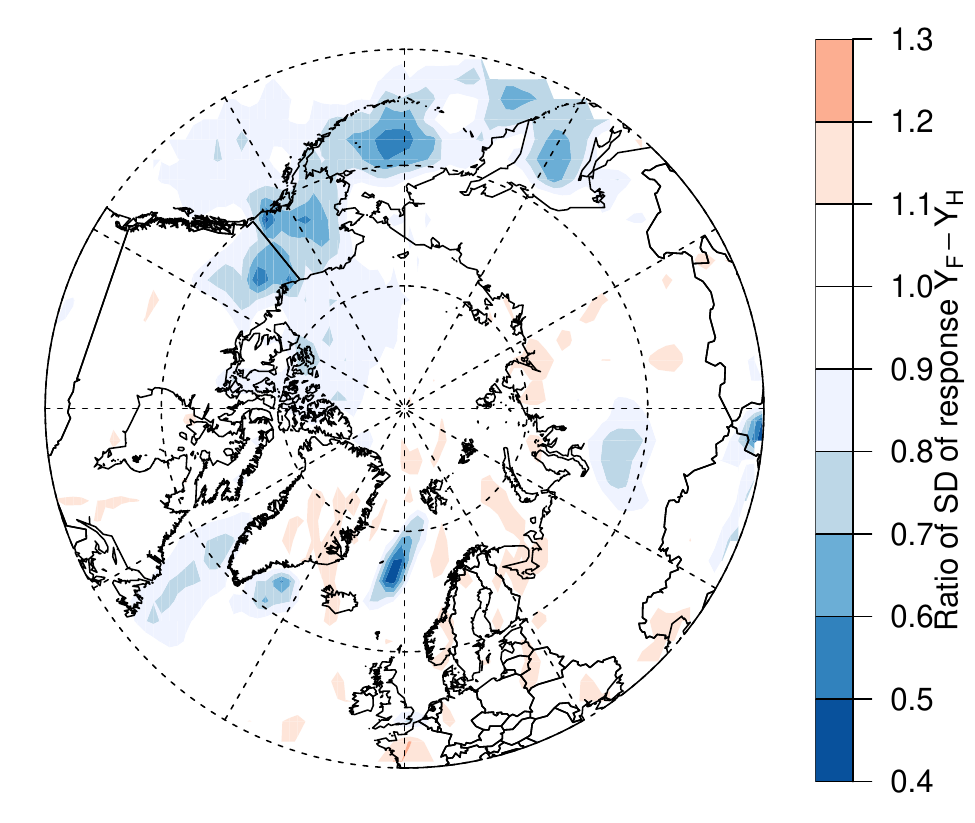}
  \end{subfigure}
  \caption{
    Effect of emergent constraints.
    The posterior mean of 
    (\subref*{fig:deltah}) 
      the historical discrepancy $Y_H - \mu_H$,
    (\subref*{fig:beta}) 
      the emergent constraint $\beta$, and
    (\subref*{fig:diff}) 
      the difference in the projected climate response $Y_H$ due to the emergent constraint $(\beta - 1) (Y_H - \mu_H)$.
    (\subref*{fig:ratio}) 
      Ratio of posterior standard deviation of the response $Y_F - Y_H$ with and without an emergent constraint.}
  \label{fig:er}
\end{figure}

Figure~\ref{fig:er} shows the effects of emergent relationships in near-surface temperature in the Arctic.
The posterior mean estimate of the historical discrepancy between the expected climate $Y_H$ and the representative climate $\mu_H$ is \SIrange{2}{3}{\celsius} across most of the Arctic (\autoref{fig:deltah}).
The historical discrepancy is largest in the Greenland and Barents seas. 
This may be due to differences in ocean heat transport simulated by the models \citep{Mahlstein2011}.
From \autoref{eqn:yt}, the expected climate response is $\Ep{Y_F - Y_H \mid Y_H}  = \mu_F - \mu_H  + (\beta - 1) (Y_H - \mu_H)$.
The difference in the projected response due to the emergent relationship is given by $(\beta - 1) (Y_H - \mu_H)$ and is plotted in \autoref{fig:diff}.
The expected warming is reduced by up to \SI{3}{\celsius} in the far north of Canada, and by around \SI{1}{\celsius} along most of the ice edge.
\autoref{fig:ratio} compares the posterior uncertainty about the climate response $Y_F - Y_H$ with and without an emergent constraint.
Around the ice edge, the emergent constraint reduces the posterior standard deviation of the climate response $Y_F - Y_H$ by \SIrange{20}{30}{\percent}.

Our posterior mean estimate of the emergent relationship in the Beaufort sea, north of Alaska in \autoref{fig:beta}, is much greater than that of \citet{Bracegirdle2013}.
Internal variability is small compared to model uncertainty in the Arctic (not shown), so the difference is not due to regression dilution in the ensemble regression estimates.
\citet{Bracegirdle2013} analyzed an ensemble of 22 CMIP5 models, some of which were excluded from the ensemble analyzed here.
Further investigation revealed that two of the models excluded from our analysis are strongly warm biased in this region, and two are strongly cold biased, but all four simulate similar climate responses.
This acts to neutralize the emergent relationship evident in the remaining models (not shown) in the analysis of \citet{Bracegirdle2013}.

\begin{figure}[!ht]
  \centering
  \begin{subfigure}[t]{0.44\textwidth}
    \subcaption{} \label{fig:svalbard}
    \includegraphics[width=\textwidth]{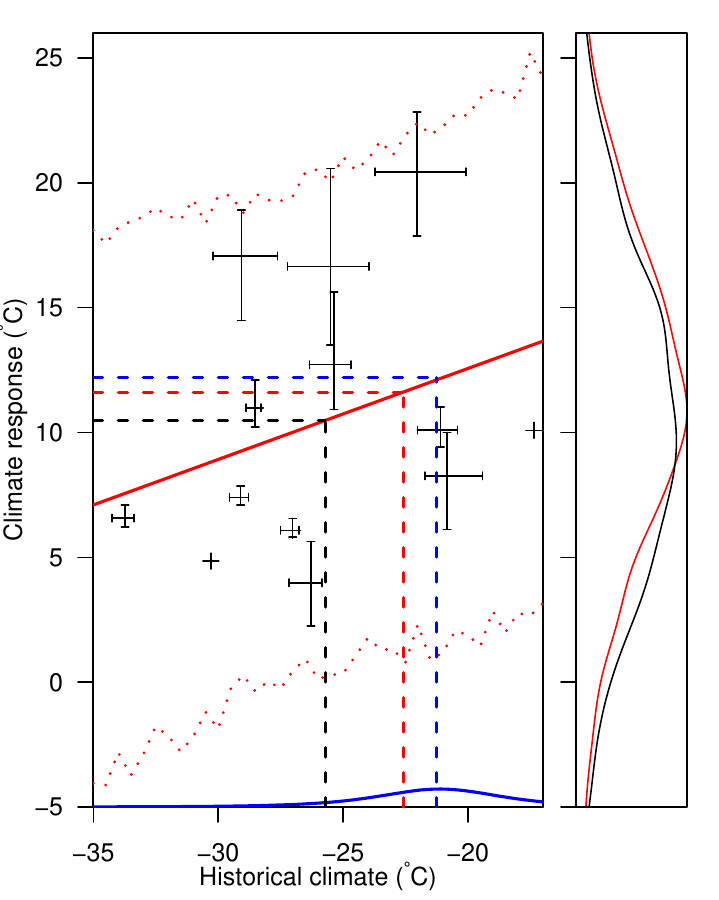}
  \end{subfigure} 
  \begin{subfigure}[t]{0.55\textwidth}
    \subcaption{} \label{fig:canada}
    \includegraphics[width=\textwidth]{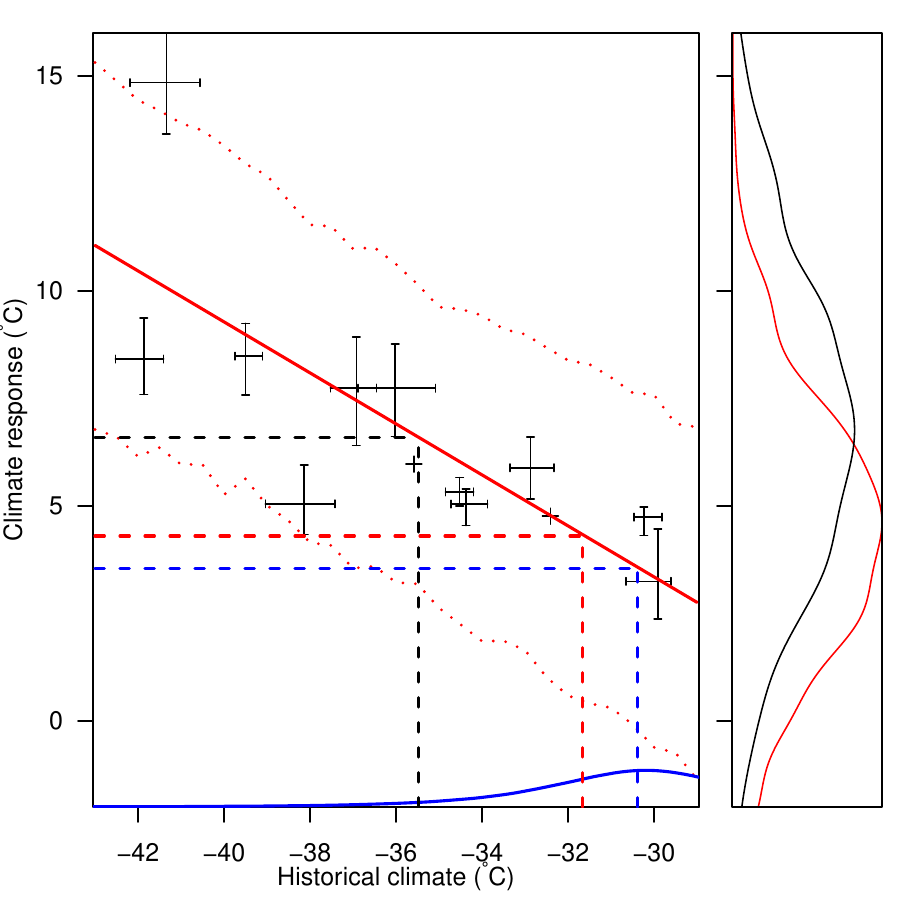}
  \end{subfigure}
   \caption{
     Gridbox details.
     Data and projections from grid boxes 
     (\subref*{fig:svalbard}) 
       north of Svalbard (\ang{81}N,\ang{39}E), and
     (\subref*{fig:canada}) 
       east of Devon Island (\ang{76}N,\ang{94}W).
     The solid red line indicates the estimated emergent relationship and the dotted red lines indicate a \SI{90}{\percent} credible interval.
     The black dashed line indicates the representative climate $\mu_H$ and climate response $\mu_F - \mu_H$.
     The red dashed line indicates the expected climate $Y_H$ and climate response $Y_F - Y_H$.
     The blue density represents the distribution of the observations.
     The blue dashed line indicates the observed climate $Z_H$ and the climate response based directly on the observations.
     Auxiliary plots in the right hand margins show the posterior distribution of the climate response $Y_F - Y_H$ with (red) and without (black) an emergent constraint.
  }
  \label{fig:details}
\end{figure}

The greatest warming occurs near the islands of Svalbard and Franz Josef Land in the north of the Barents sea.
\autoref{fig:svalbard} investigates the strong warming near Svalbard in detail.
The representative climate response $\mu_F - \mu_H$ in \autoref{fig:svalbard} is already high at \SI{10.5}{\celsius} (\SI{90}{\percent} equal-tailed credible interval \SIrange{7.7}{13.3}{\celsius}).
The representative response may be influenced by 3 models with unusually large responses.
There is a positive emergent relationship $\beta = 1.4$ (\numrange[range-phrase={,}]{0.8}{2.0}) at this grid box, and a historical discrepancy of $Y_H - \mu_H = \SI{3.0}{\celsius}$ (\SIrange[range-phrase={,}]{-2.7}{8.2}{\celsius}).
The emergent relationship predicts an additional \SI{1.1}{\celsius} (\SIrange[range-phrase={,}]{-1.6}{4.8}{\celsius}) of warming.
This is relatively insignificant compared to the uncertainty about the response, even when conditioned on the historical climate.
The emergent relationship here does little to constrain our uncertainty about the climate response.

\citet{Bracegirdle2013} also estimated a positive emergent relationship over Svalbard, Franz Josef Land and parts of Siberia, similar to that in \autoref{fig:beta}.
The posterior probability that $\beta > 1$ exceeds 0.90 over Western Siberia.
Emergent constraints in air temperatures over polar land regions are particularly relevant for constraining estimates of changes in permafrost, which by melting in future could lead to accelerated emissions in greenhouse gases such as methane \citep{Burke2013}. 
There are significant differences in model temperatures over polar land regions related to model representation of processes such as snow physics and soil hydrology \citep{Koven2013,Slater2013}.
It remains an interesting open question as to why models are showing a positive emergent relationship in the vicinity Western Siberia. 

In contrast, a negative emergent relationship is visible in the North West Passage near Devon Island in northern Canada in \autoref{fig:canada}.
The representative climate response $\mu_F - \mu_H$ in \autoref{fig:canada} is more moderate at \SI{6.6}{\celsius} (\SIrange[range-phrase={,}]{5.1}{8.0}{\celsius}).
There is a negative emergent relationship $\beta = 0.4$ (\numrange[range-phrase={,}]{0.2}{0.7}) and a historical discrepancy of $Y_H - \mu_H = \SI{3.7}{\celsius}$ (\SIrange[range-phrase={,}]{-1.1}{8.4}{\celsius}).
The emergent relationship combines with the historical discrepancy to project \SI{2.2}{\celsius} (\SIrange[range-phrase={,}]{-5.3}{0.6}{\celsius}) less warming than the representative model.
At this grid box, our uncertainty is usefully constrained by the emergent relationship.
The modification to both the mean and standard deviation of the posterior projected response is shown in the right hand margin of \autoref{fig:canada}.
The posterior standard deviation of the projected response $Y_F - Y_H$ is reduced by \SI{18}{\percent}, falling from \SI{3.9}{\celsius} to \SI{3.2}{\celsius}.

The examples of Svalbard and Devon Island in \autoref{fig:details} both demonstrate the important role of observation and sampling uncertainty when combining models and observations.
Due to the sparsity of observations in these remote regions, the observation uncertainty is quite large relative to the model uncertainty.
In both cases, there is noticeable shrinkage of the posterior mean estimate of the historical climate $Y_H$ away from the observations $Z_H$ and towards the representative climate $\mu_H$.
As a result, the projected response $Y_F - Y_H$ lies closer to the representative response $\mu_F - \mu_H$ than it would if observation uncertainty were ignored.

\section{Conclusion}
\label{sec:conclusion}

Emergent relationships have become an important topic in climate science for their potential to constrain our uncertainty about future climate change.
In this study, we have argued that such relationships can be used to constrain discrepancies due to model inadequacy, if a physical mechanism for the relationship can be identified. 
The negative emergent constraint on near surface temperature in the Arctic is well understood, and our analysis broadly confirms the findings of previous studies.
The projected warming in the Arctic is reduced by up to \SI{3}{\celsius} by the emergent constraint.
Internal variability in the Arctic is large compared to lower latitudes, but is dwarfed by model uncertainty due to the difficulty of representing the many complex processes involved in simulating sea ice, snow cover and the polar vortex.
Therefore, regression dilution is unlikely to have significantly biased previous studies of Arctic climate change.
However, the sparsity of observations in the Arctic means there is significant observation uncertainty, and this is the first time that observation uncertainty has been accounted for when exploiting emergent constraints.
Shrinkage of the expected climate towards the representative climate results in differences of up to \SI{1}{\celsius} in the projected response compared to estimates based on the observations directly.

The main contribution of this study is to link the concepts of model inadequacy in an ensemble of models and emergent relationships.
The proposed Bayesian hierarchical framework also allows the inclusion of multiple runs from each simulator for the first time in a practical application.
This allows us to separate uncertainty due to differences between models from internal variability within models.
It is differences in the representation of key processes that lead to emergent relationships.
Initial conditions should be forgotten over sufficiently long time scales, and therefore should not lead to emergent behavior.
We have shown that if internal variability is not accounted for, then projections based on emergent constraints may be biased.
Future multi-model studies exploiting emergent constraints should include multiple runs from each simulator in order to separate model uncertainty from internal variability and avoid potentially biased projections.
Another unique aspect of the framework proposed here is the separation of natural variability and observation uncertainty in the climate system.

The framework proposed in this study allows robust estimation and projection using emergent constraints, but there are still open problems to be addressed both in general multi-model experiments and emergent relationships.
The methodology proposed here allows projection of time mean climate accounting for uncertainty due to natural variability.
If time-series realizations of natural variability are required within the future study period, e.g., for adaptation studies, then our methodology could be extended using the time-series approach proposed by \citet{Tebaldi2009}, or by transforming observations as proposed by \citet{Poppick2016}.
Where emergent constraints have been studied at a local level, rather than an aggregate or process level, they have been analyzed one grid box at a time.
Ignoring spatial dependence between grid boxes may lead to overly smooth estimates, due to differences in feature placement between models.
In order to obtain physically realistic inferences, spatial statistical methods are required that can represent spatial dependence while accounting for differences between models.

The CMIP5 multi-model ensemble included four future emissions scenarios, but we have analyzed only one.
Like climate models, emissions scenarios are difficult to interpret together as an ensemble.
Innovative methods are required to extract meaningful probabilistic projections that span the likely range of future emissions.
Another source of uncertainty not usually addressed in multi-model experiments is uncertainty about the internal parameters of the climate models.
The computational cost of running large perturbed-parameter ensembles is prohibitive.
However, each model undergoes a tuning process during which the internal parameters are tested and fixed \citep{Hourdinetal2017}.
Statistical emulators for key quantities, trained during this tuning process, might provide a way of integrating parameter uncertainty into multi-model experiments to provide a more holistic assessment of our uncertainty.

In order to satisfy the assumption of exchangeability we analyze only a subset of the available models.
By adopting this approach we risk losing valuable information contained in runs from other models and ignoring more detailed insights about model dependence that could be gained by comparing model outputs.
In principle, additional levels could be added to the hierarchy proposed here to represent models that share components or were built by the same group.
However, the complex overlapping relationships make such a highly structured approach problematic.
Current methods for quantifying model dependence based on comparing spatial-temporal output patterns ignore all the prior knowledge we have about have about the relationships between models.
One way forward might be to develop frameworks that combine grouping based on comparing spatial-temporal outputs with simple judgments based on prior knowledge of model inter-dependence.
Until alternative methods are found, we recommend thinning the ensemble to obtain an approximately exchangeable set of models and transparently documenting the thinning process.
This does require some prior knowledge on the part of the analyst.
However, the burden could be alleviated by establishing standard lists of models, e.g., centers submitting to model inter-comparison projects could be asked to nominate a primary model for analysis.
This opens up the interesting question of multi-model experiment design.
However, the greatest statistical challenge in climate projection is meaningful quantification of model inadequacy.
The results here and in \citet{Rougieretal2013} demonstrate how far we can go with simple judgments.
However, additional co-operation between statisticians and climate scientists is required to make further progress.

\bibliographystyle{plainnat}
\bibliography{library}

\end{document}


\begin{center}
{\large\bf SUPPLEMENTARY MATERIAL}
\end{center}

\section*{Derivation of Gibbs'-Metropolis updating equations}

For the purposes of computation it is more convenient to work with precisions than variances, so let
\begin{align*}
  \tau_m = 1 / \sigma_m^2  \ \text{for} \ m = 1,\ldots,M; \quad 
  \phi_m = 1 / \varphi_m^2 \ \text{for} \ m = 1,\ldots,M
\end{align*}
\vspace{-2\baselineskip}
\begin{align*}
  \tau_H          & = 1 / \sigma_H^2         ; & 
  \tau_{F \mid H} & = 1 / \sigma_{F \mid H}^2; &
  \tau_a          & = 1 / \sigma_a^2         ; & 
  \phi_a          & = 1 / \varphi_a^2 \\
  \tau_{\disc_H}  & = 1 / \sigma_{\disc_H}^2 ; & 
  \tau_{\disc_F}  & = 1 / \sigma_{\disc_F}^2 ; &
  \tau_W          & = 1 / \sigma_W^2         ; & 
  \tau_{\disc_W}  & = 1 / \sigma_{\disc_W}^2.
\end{align*}
The complete model defined by Equations~1--8 and 9--11 can be rewritten as
\begin{align*}
  X_{Hmr} \mid X_{Hm} & \sim \Np{X_{Hm}}{\tau_m^{-1}}  &
  X_{Fmr} \mid X_{Fm} & \sim \Np{X_{Fm}}{\left( \phi_m \tau_m \right)^{-1}} \\
  X_{Hm}              & \sim \Np{\mu_H }{\tau_H^{-1}}  &
  X_{Fm} \mid X_{Hm}  & \sim \Np{\mu_F + \beta \left( X_{Hm} - \mu_H \right)}{\tau_{F \mid H}^{-1}} \\
  \tau_m & \sim \Gap{\frac{\nu_H}{2}}{\frac{\nu_H \esigmahm}{2}} &
  \phi_m & \sim \Gap{\frac{\nu_F}{2}}{\frac{\nu_F \esigmafm}{2}} \\
  Y_{Ha} \mid Y_H & \sim \Np{Y_H}{\tau_a^{-1}}    &
  Y_{Fa} \mid Y_F & \sim \Np{Y_F}{\left( \phi_a \tau_a \right)^{-1}}  \\
  Y_H          & \sim \Np{\mu_H}{\tau_{\disc_H}^{-1}}  &
  Y_F \mid Y_H & \sim \Np{\mu_F + \beta \left( Y_H - \mu_H \right)}
                         {\tau_{\disc_F}^{-1}} \\
  \tau_a & \sim \Gap{\frac{\nu_{Ha}}{2}}{\frac{\nu_{Ha} \esigmahm}{2}} &
  \phi_a & \sim \Gap{\frac{\nu_{Fa}}{2}}{\frac{\nu_{Fa} \esigmafm}{2}} \\
  W_i    & \sim \Np{\mu_W }{\tau_W^{-1}} &   
  \mu_W  & \sim \Np{Y_{Ha}}{\tau_{\disc_W}^{-1}}
\end{align*}
where
\begin{align*}
  \tau_{\disc_H}          & = \tau_H          / \kappa^2   &
  \tau_{\disc_{F \mid H}} & = \tau_{F \mid H} / \kappa^2   &
  \tau_{\disc_W}          & = \tau_W          / \kappa_W^2 \\
  \nu_{Ha} & = \nu_H / \kappa^2  &
  \nu_{Fa} & = \nu_F / \kappa^2.  &
\end{align*}

Let $\bX = (X_{tmr}, s \in \lbrace H,F \rbrace, m = 1,\ldots,M, r = 1,\ldots,R_{tm})^\prime$ be the model outputs, $\bY = ( Y_H, Y_{Ha}, \tau_a)^\prime$ be the latent state of the climate system, $\btheta = (\mu_H, \mu_F, \beta, \tau_H, \tau_{F \mid H}, \psi^2, \phi^2, \nu_H, \nu_F)^\prime$ be the ensemble parameters, $\bchi = (X_{Hm}, X_{Fm}, \tau_m, \phi_m, m = 1,\ldots,M)^\prime$ be the latent model states, $\bW = (W_i, i = 1,\ldots,N)$ be the reanalysis outputs, and $\bomega = (\mu_W, \tau_W)^\prime$ be the reanalysis parameters.
The future state of the climate system defined by $Y_F$, $Y_{Fa}$ and $\phi_a$ are purely predictive quantities and can be sampled after sampling of all other quantities is complete, using the equations above.

The joint posterior can be decomposed as
\begin{align*}
  \Prp{\bY,\bchi,\btheta,\bomega \mid \bX, \bW} 
    & \propto \Prp{\bW \mid \bomega} \Prp{\bY \mid \btheta, \bomega}
              \Prp{\bX \mid \bchi} \Prp{\bchi \mid \btheta} 
              \Prp{\btheta} \Prp{\bomega}
\end{align*}
The likelihood of the reanalysis outputs $\bW$ given the reanalysis parameters $\bomega$ is proportional to
\begin{align*}
  \Prp{\bW \mid \bomega} 
    & \propto \prod_{i=1}^N \tau_W^{1/2} \exp \left( - \frac{\tau_W}{2} 
        \left( W_i - \mu_W \right)^2 \right).
\end{align*}
The likelihood of the system $\bY$ given the ensemble parameters $\btheta$ and the reanalysis parameters $\bomega$ is proportional to
\begin{align*}
  \Prp{\bY \mid \btheta, \bomega}
    & \propto \tau_{\disc_H}^{1/2} \exp \left( - \frac{\tau_{\disc_H}}{2}
                \left( Y_H - \mu_H \right)^2 \right)
              \tau_a^{1/2} \exp \left( - \frac{\tau_a}{2}
                \left( Y_{Ha} - Y_H \right)^2 \right)  \\
    & \quad   \frac{\left( \frac{\nu_{Ha} \esigmahm}{2} \right)^{\nu_{Ha} / 2}}
                     {\Gamma \left( \nu_{Ha} / 2 \right)}
                \tau_a^{\nu_{Ha} / 2 - 1}
                \exp \left( - \frac{\nu_{Ha} \esigmahm}{2} \tau_a \right).
\end{align*}
The likelihood of the model outputs $\bX$ given the latent model states $\bchi$ is proportional to
\begin{align*}
  \Prp{\bX \mid \bchi} 
    & \propto \prod_{m=1}^M \prod_{r=1}^{R_{Hm}} 
                \tau_m^{1/2} \exp \left( - \frac{\tau_m}{2} 
                  \left( X_{Hmr} - X_{Hm} \right)^2 \right)   \\
    & \quad   \prod_{m=1}^M \prod_{r=1}^{R_{Fm}} 
                \left( \phi_m \tau_m \right)^{1/2} 
                  \exp \left( - \frac{\phi_m \tau_m}{2} 
                    \left( X_{Fmr} - X_{Fm} \right)^2 \right).
\end{align*}
The likelihood of the model states $\bchi$ given the ensemble parameters $\theta$ is proportional to
\begin{align*}
  \Prp{\bchi \mid \btheta} 
    & \propto \prod_{m=1}^M \tau_H^{1/2} 
                \exp \left( - \frac{\tau_H}{2} 
                  \left( X_{Hm} - \mu_H \right)^2 \right) \\
    & \quad   \prod_{m=1}^M \tau_F^{1/2} 
                \exp \left( - \frac{\tau_F}{2} 
                  \left( X_{Fm} - \mu_F - \beta \left( X_{Hm} - \mu_H \right)
                    \right)^2 \right) \\
    & \quad   \prod_{m=1}^M
                \frac{\left( \frac{\nu_H \esigmahm     }{2} \right)^{\nu_H / 2}}
                     {\Gamma \left( \nu_H / 2 \right)}
                \tau_m^{\nu_H / 2 - 1}
                \exp \left( - \frac{\nu_H \esigmahm     }{2} \tau_m \right) \\
    & \quad   \prod_{m=1}^M 
                \frac{\left( \frac{\nu_F \esigmafm}{2} \right)^{\nu_F / 2}}
                     {\Gamma \left( \nu_F / 2 \right)}
                \phi_m^{\nu_F / 2 - 1}
                \exp \left( - \frac{\nu_F \esigmafm}{2} \phi_m \right).
\end{align*}
The joint prior distribution of the ensemble parameters $\btheta$ is proportional to
\begin{align*}
  \Prp{\btheta} 
    & \propto \exp \left( - \frac{b_{\mu_H}}{2} 
                \left( \mu_H   - a_{\mu_H} \right)^2 \right)
              \exp \left( - \frac{b_{\mu_F}}{2} 
                \left( \mu_F   -    \mu_H \right)^2 \right)
              \exp \left( - \frac{b_\beta}{2} 
                \left( \beta - a_\beta \right)^2 \right)  \\
    & \quad   \tau_H^{a_{\tau_H} - 1} 
                \exp \left( - b_{\tau_H} \tau_H \right)
              \tau_F^{a_{\tau_F} - 1} 
                \exp \left( - b_{\tau_F} \tau_F \right)
              \nu_H^{a_{\nu_H} - 1} 
                \exp \left( - b_{\nu_H} \nu_H \right)
              \nu_F^{a_{\nu_F} - 1} 
                \exp \left( - b_{\nu_F} \nu_F \right)         \\
    & \quad   {\left( \esigmahm      \right)}^{a_{\esigmahm     } - 1} 
                \exp \left( - b_{\esigmahm     } \esigmahm      \right)
              {\left( \esigmafm \right)}^{a_{\esigmafm} - 1} 
                \exp \left( - b_{\esigmafm} \esigmafm \right).
\end{align*}
The joint prior distribution of the reanalysis parameters $\bomega$ is proportional to
\begin{align*}
  \Prp{\bomega} 
    & \propto \tau_{\disc_W}^{1/2} \exp \left( - \frac{\tau_{\disc_W}}{2} 
                \left( \mu_W - Y_{Ha} \right)^2 \right)
              \tau_W^{a_{\tau_W} - 1} 
                \exp \left( - b_{\tau_W} \tau_W \right).
\end{align*}

The full conditional distributions of the system quantities $\bY$ are
\begin{align*}
  Y_{Ha} \mid \ldots & \sim
    \Np{\frac{\tau_a Y_H + \tau_{\disc_W} \mu_W}{\tau_a + \tau_{\disc_W}}}
       {\left( \tau_a + \tau_{\disc_W} \right)^{-1}} \\
  Y_H \mid \ldots & \sim
    \Np{\frac{\tau_{\disc_H} \mu_H + \tau_a Y_{Ha}}
             {\tau_{\disc_H} + \tau_a}}
       {\left( \tau_{\disc_H} + \tau_a \right)^{-1}} \\
  \tau_a \mid \ldots & \sim
    \Gap{\frac{\nu_{Ha} + 1}{2}}
        {\frac{\nu_{Ha} \esigmahm + \left( Y_{Ha} - Y_H \right)^2}{2}} \\
\end{align*}
The full conditional distributions of the reanalysis parameters $\bomega$ are
\begin{align*}
  \mu_W \mid \ldots & \sim
    \Np{\frac{\tau_W \sum_i W_i + \tau_{\disc_W} Y_{Ha}}
             {\tau_W N + \tau_{\disc_W}}}
       {\left( \tau_W N + \tau_{\disc_W} \right)^{-1}} \\
  \tau_W \mid \ldots & \sim
    \Gap{a_{\tau_W} + \frac{N + 1}{2}}
        {b_{\tau_W} + \frac{1}{2} \sum_i \left( W_i - \mu_W \right)^2 +
                      \frac{1}{2} \kappa_W^{-2} \left( \mu_W - Y_{Ha} \right)^2}
\end{align*}
The full conditional distributions of the latent model states $\bchi$ are
\begin{align*}
  X_{Fm} \mid \ldots & \sim
    \Np{\frac{\tau_F \left( \mu_F + \beta \left( X_{Hm} - \mu_H \right) \right)
                + \phi_m \tau_m \sum_r X_{Fmr}}
             {\tau_F + \phi_m \tau_m R_{Fm}}}
       {\left( \tau_F + \phi_m \tau_m R_{Fm} \right)^{-1}} \\
  X_{Hm} \mid \ldots & \sim
    \Np{\frac{\tau_H \mu_H
                + \tau_F \beta \left( X_{Fm} - \mu_F + \beta \mu_H \right)
                + \tau_m \sum_r X_{Hmr}}
             {\tau_H + \tau_F \beta^2 + \tau_m R_{Hm}}}
       {\left( \tau_H + \tau_F \beta^2 + \tau_m R_{Hm} \right)^{-1}} \\
  \tau_m \mid \ldots & \sim
    \Gap{\frac{\nu_H + N_{Hm} + N_{Fm}}{2}}
        {\frac{\nu_H \esigmahm + \sum_r \left( X_{Hmr} - X_{Hm} \right)^2 
                 + \phi_m \sum_r \left( X_{Fmr} - X_{Fm} \right)^2}{2}} \\
  \phi_m \mid \ldots & \sim
    \Gap{\frac{\nu_F + N_{Fm}}{2}}
        {\frac{\nu_F \esigmafm 
                 + \tau_m \sum_r \left( X_{Fmr} - X_{Fm} \right)^2}{2}}
\end{align*}
The full conditional distributions of the ensemble parameters $\btheta$ are
\begin{align*}
  \mu_H \mid \ldots & \sim
    \Np{\tilde{\mu}_H}
       {\left( b_{\mu_H} + b_{\mu_F} + \tau_H M + \tau_F \beta^2 M + 
          \tau_{\disc_H} \right)^{-1}} \\
  \mu_F \mid \ldots & \sim
    \Np{\frac{b_{\mu_F} \mu_H 
          + \tau_F \sum_m 
              \left( X_{Fm} - \beta \left( X_{Hm} - \mu_H \right) \right)}
             {b_{\mu_F} + \tau_F M}}
       {\left( b_{\mu_F} + \tau_F M \right)^{-1}} \\
  \beta \mid \ldots & \sim
    \Np{\frac{b_\beta a_\beta + 
              + \tau_F \sum_m \left( X_{Hm} - \mu_H \right)
                              \left( X_{Fm} - \mu_F \right)}
             {b_\beta + \tau_F \sum_m \left( X_{Hm} - \mu_H \right)^2}}
       {\left( b_\beta + \tau_F \sum_m \left( X_{Hm} - \mu_H \right)^2 \right)^{-1}} \\
  \tau_H \mid \ldots & \sim
    \Gap{a_{\tau_H} + \frac{M + 1}{2}}
        {b_{\tau_H} + 
         \frac{\sum_m \left( X_{Hm} - \mu_H \right)^2 
                   + \kappa^{-2} \left( Y_H - \mu_H\right)^2}{2}} \\
  \tau_F \mid \ldots & \sim
    \Gap{a_{\tau_F} + \frac{M}{2}}
        {b_{\tau_F} + 
         \frac{\sum_m \left( X_{Fm} - \mu_F - 
                         \beta \left( X_{Hm} - \mu_H \right) \right)^2}{2}} \\
  \esigmahm \mid \ldots & \sim 
    \Gap{a_{\esigmahm} + \frac{\nu_H M + \nu_{Ha}}{2}}
        {b_{\esigmahm } + \frac{\nu_H \sum_m \tau_m + \nu_{Ha} \tau_a}{2}} \\
  \esigmafm \mid \ldots & \sim 
    \Gap{a_{\esigmafm} + \frac{\nu_F M}{2}}
        {b_{\esigmafm}  + \frac{\nu_F \sum_m \phi_m}{2}}
\end{align*}
where
\begin{align*}
  \tilde{\mu}_H  & = \frac{b_{\mu_H} a_{\mu_H} 
          + b_{\mu_F} \mu_F + \tau_H \sum_m X_{Hm}
          - \tau_F \beta \sum_m \left( X_{Fm} - \mu_F - \beta X_{Hm} \right)
          + \tau_{\disc_H} Y_H}
           {b_{\mu_H} + b_{\mu_F} + \tau_H M + \tau_F \beta^2 M + 
              \tau_{\disc_H}}.
\end{align*}

\newpage

The full conditional distributions of the degrees-of-freedom $\nu_H$ and $\nu_F$ do not correspond to any standard distribution.
The likelihoods associated with $\nu_H$ and $\nu_F$ are
\begin{align*}
  l \left( \nu_H \right) =
            \frac{\beta_{Ha}^{\alpha_{Ha}}}
                 {\Gamma \left( \alpha_{Ha} \right)} 
            \tau_a^{\alpha_{Ha} - 1} \exp \left( - \beta_{Ha} \tau_a \right)
    \prod_m \frac{\beta_H^{\alpha_H}}
                 {\Gamma \left( \alpha_H \right)} 
            \tau_m^{\alpha_H - 1} \exp \left( - \beta_H \tau_m \right)
\end{align*}
and
\begin{align*}
  l \left( \nu_F \right) = 
     \prod_m \frac{\beta_F^{\alpha_F}}
                 {\Gamma \left( \alpha_F \right)} 
            \phi_m^{\alpha_F - 1} \exp \left( - \beta_F \phi_m \right)
\end{align*}
where
\begin{align*}
  \alpha_{Ha} & = \nu_{Ha} / 2,  & \beta_{Ha} & = \nu_{Ha} \esigmahm / 2,  &
  \alpha_H    & = \nu_H    / 2,  & \beta_H    & = \nu_H    \esigmahm / 2,  &
  \alpha_F    & = \nu_F    / 2,  & \beta_F    & = \nu_F    \esigmafm / 2.
\end{align*}
The prior densities of $\nu_H$ and $\nu_F$ are
\begin{align*}
  p \left( \nu_H \right) \propto 
      \nu_H^{a_{\nu_H} - 1} \exp \left( - b_{\nu_H} \nu_H \right)
  \quad \text{and} \quad
  p \left( \nu_F \right) \propto
      \nu_F^{a_{\nu_F} - 1} \exp \left( - b_{\nu_F} \nu_F \right).
\end{align*}

The posterior distributions of $\nu_H$ and $\nu_F$ conditional on the current state of the other parameters can be sampled using the Metropolis-Hastings algorithm.
For each $s \in \lbrace H, F \rbrace$:
\begin{enumerate}
  \item Sample a new state $\nu_t^\star$ from $q \left( \nu_t^\star \mid \nu_t \right)$;
  \item Calculate the Hastings ratio
    \begin{equation*}
      r \left( \nu_t^\star, \nu_t \right)
        = \frac{l(\nu_t^\star) p(\nu_t^\star) q(\nu_t^\star \mid \nu_t)}
               {l(\nu_t) p(\nu_t) q(\nu_t \mid \nu_t^\star)};
    \end{equation*}
  \item Accept the new state $\nu_t^\star$ with probability
    \begin{equation*}
      a(\nu_t^\star, \nu_t) 
        = \min (1, r \left( \nu_t^\star, \nu_t \right)).
    \end{equation*}
\end{enumerate}
where $q(\nu_t^\star \mid \nu_t) = \Gap{\nu_t \lambda_t}{\lambda_t}$ is the proposal distribution, with expectation $\nu_t$ and variance controlled by the free parameter $\lambda_t$.
The acceptance rate of the Metropolis step can be controlled using the parameter $\lambda_t$.

\newpage
\section*{Ensemble thinning}

An extended version of the CMIP5 surface temperature data analyzed by 
\citet{Bracegirdle2013} was considered for analysis. 
The mean climates over 30 winters (December-January-February) are compared between December 1975 and January 2005 from the historical scenario, and between December 2069 and January 2099 from the RCP4.5 scenario. 
The five year shift in the historical period compared to \citet{Bracegirdle2013} provides slightly better compatibility with the latest observation and reanalysis data sets.
Several of these data sets begin in 1979 when satellite observations become
prevalent. 
A total of 216 runs from 37 CMIP5 models were included in the full ensemble, 128 runs of the historical scenario and 88 of the RCP4.5 scenario.
The complete list of models and details of their major components are given in \autoref{tab:coex_models}.

\begin{sidewaystable}[!p]
  \begin{center}
  \begin{tiny}
  \begin{tabular}{llllllll}
    \hline \hline
    \centering 
    Modelling centre & Model & Atmosphere & Atmosphere res. &
    Ocean & Ocean res. & Sea ice & Land surface \\
    \hline
CSIRO-BOM & ACCESS1.0 & HadGEM2 (r1.1) & 1.9 x 1.9 L38 & MOM4.1 & 1.0 x 1.0 L50 & CICE4.1 & MOSES2.2 \\
CSIRO-BOM & ACCESS1.3 & UKMO GA 1.0 & 1.9 x 1.9 L38 & MOM4.1 & 1.0 x 1.0 L50 & CICE4.1 & CABLE \\
BCC & BCC-CSM1.1 & BCC\_AGCM2.1 & 2.8 x 2.8 L26 & MOM4-L40 & 1.0 x 1.0 L40 & GFDL SIS & BCC-AVIM1.0 \\
\color{red}{BCC} & \color{red}{BCC-CSM1.1(m)} & \color{red}{BCC\_AGCM2.1} & \color{red}{1.1 x 1.1 L26} & \color{red}{MOM4-L40} & \color{red}{1.0 x 1.0 L40} & \color{red}{GFDL SIS} & \color{red}{BCC-AVIM1.0} \\
GCESS & BNU-ESM & CAM3.5 & 2.8 x 2.8 L26 & MOM4.1 & 1.0 x 1.0 L50 & CICE4.1 & CoLM+BNUDGVM \\
\color{red}{CCCMA} & \color{red}{CanESM2} & \color{red}{CanAM4} & \color{red}{2.8 x 2.8 L35} & \color{red}{CanOM4} & \color{red}{1.4 x 1.4 L40} & \color{red}{Included} & \color{red}{CLASS 2.7; CTEM} \\
NCAR & CCSM4 & CAM4 & 1.2 x 1.2 L27 & POP2 & 1.0 x 1.0 L60 & CICE4 & CLM4 \\
NSF-DOE-NCAR & CESM1(BGC) & CAM4 & 1.2 x 1.2 L27 & POP2 & 1.0 x 1.0 L60 & CICE4 & CLM4 \\
\color{red}{NSF-DOE-NCAR} & \color{red}{CESM1(CAM5)} & \color{red}{CAM5} & \color{red}{1.2 x 1.2 L27} & \color{red}{POP2} & \color{red}{1.0 x 1.0 L60} & \color{red}{CICE4} & \color{red}{CLM4} \\
NSF-DOE-NCAR & CESM1(WACCM) & WACCM4 & 2.5 x 2.5 L66 & POP2 & 1.0 x 1.0 L60 & CICE4 & CLM4 \\
CMCC & CMCC-CM & ECHAM5 & 0.8 x 0.8 L31 & OPA8.2 & 2.0 x 2.0 L31 & LIM2 & N / A \\
CMCC & CMCC-CMS & ECHAM5 & 1.9 x 1.9 L95 & OPA8.2 & 2.0 x 2.0 L31 & LIM2 & N / A \\
CNRM-CERFACS & CNRM-CM5 & ARPEGE-Climat & 1.4 x 1.4 L31 & NEMO & 0.7 x 0.7 L42 & Gelato5 & SURFEX \\
CSIRO-QCCCE & CSIRO-Mk3.6.0 & Included & 1.9 x 1.9 L18 & MOM2.2 & 1.9 x 1.9 L31 & Included & Included \\
\color{red}{EC-EARTH} & \color{red}{EC-EARTH} & \color{red}{IFS c3 1r1} & \color{red}{1.1 x 1.1 L62} & \color{red}{NEMO\_ecmwf} & \color{red}{1.0 x 1.0 L31} & \color{red}{LIM2} & \color{red}{HTESSEL} \\
\color{red}{LASG-CESS} & \color{red}{FGOALS-g2} & \color{red}{GAMIL2} & \color{red}{2.8 x 2.8 L26} & \color{red}{LICOM2} & \color{red}{1.0 x 1.0 L30} & \color{red}{CICE4-LASG} & \color{red}{CLM3} \\
FIO & FIO-ESM & CAM3.0 & 2.8 x 2.8 L26 & POP2 & 1.0 x 1.0 L40 & CICE4 & CLM3.5 \\
NOAA GFDL & GFDL-CM3 & Included & 2.5 x 2.5 L48 & MOM4.1 & 1.0 x 1.0 L50 & SIS & Included \\
\color{red}{NOAA GFDL} & \color{red}{GFDL-ESM2G} & \color{red}{AM2.1} & \color{red}{2.5 x 2.5 L24} & \color{red}{GOLD} & \color{red}{1.0 x 1.0 L63} & \color{red}{SIS} & \color{red}{Included} \\
NOAA GFDL & GFDL-ESM2M & AM2.1 & 2.5 x 2.5 L60 & MOM4.1 & 1.0 x 1.0 L50 & SIS & Included \\
NASA GISS & GISS-E2-H & Inlcuded & 2.5 x 2.5 L40 & HYCOM & 1.0 x 1.0 L26 & Included & Included \\
\color{red}{NASA GISS} & \color{red}{GISS-E2-R} & \color{red}{Inlcuded} & \color{red}{2.5 x 2.5 L40} & \color{red}{Russell} & \color{red}{1.0 x 1.0 L32} & \color{red}{Included} & \color{red}{Included} \\
NIMR/KMA & HadGEM2-AO & HadGAM2 & 1.9 x 1.9 L60 & Included & 1.9 x 1.9 & Included & Included \\
MOHC & HadGEM2-CC & HadGAM2 & 1.9 x 1.9 L60 & Included & 1.9 x 1.9 & Included & Included \\
\color{red}{MOHC} & \color{red}{HadGEM2-ES} & \color{red}{HadGAM2} & \color{red}{1.9 x 1.9 L38} & \color{red}{Included} & \color{red}{1.0 x 1.0 L40} & \color{red}{Included} & \color{red}{Included} \\
\color{red}{INM} & \color{red}{INM-CM4} & \color{red}{Included} & \color{red}{2.0 x 2.0 L21} & \color{red}{Included} & \color{red}{1.0 x 1.0 L40} & \color{red}{Included} & \color{red}{Included} \\
IPSL & IPSL-CM5A-LR & LMDZ5 & 3.8 x 3.8 L39 & NEMO & 2.0 x 2.0 L31 & Included & Included \\
\color{red}{IPSL} & \color{red}{IPSL-CM5A-MR} & \color{red}{LMDZ5} & \color{red}{2.5 x 2.5 L39} & \color{red}{NEMO} & \color{red}{2.0 x 2.0 L31} & \color{red}{Included} & \color{red}{Included} \\
IPSL & IPSL-CM5B-LR & LMDZ5 & 3.8 x 3.8 L39 & NEMO & 2.0 x 2.0 L31 & Included & Included \\
\color{red}{MIROC} & \color{red}{MIROC5} & \color{red}{MIROC-AGCM6} & \color{red}{1.4 x 1.4 L40} & \color{red}{COCO4.5} & \color{red}{1.4 x 1.4 L50} & \color{red}{Included} & \color{red}{MATSIRO} \\
MIROC & MIROC-ESM & MIROC-AGCM & 2.8 x 2.8 L80 & COCO3.4 & 1.4 x 1.4 L44 & Included & MATSIRO \\
MIROC & MIROC-ESM-CHEM & MIROC-AGCM & 2.8 x 2.8 L80 & COCO3.4 & 1.4 x 1.4 L44 & Included & MATSIRO \\
\color{red}{MPI-M} & \color{red}{MPI-ESM-LR} & \color{red}{ECHAM6} & \color{red}{1.9 x 1.9 L47} & \color{red}{MPIOM} & \color{red}{1.5 x 1.5 L40} & \color{red}{Included} & \color{red}{JSBACH} \\
MPI-M & MPI-ESM-MR & ECHAM6 & 1.9 x 1.9 L95 & MPIOM & 0.4 x 0.4 L40 & Included & JSBACH \\
\color{red}{MRI} & \color{red}{MRI-CGCM3} & \color{red}{MRI-AGCM3.3} & \color{red}{1.1 x 1.1 L48} & \color{red}{MRI.COM3} & \color{red}{1.0 x 1.0 L50} & \color{red}{MRI.COM3} & \color{red}{HAL} \\
NCC & NorESM1-M & CAM4-Oslo & 2.5 x 2.5 L26 & NorESM-Ocean & 1.1 x 1.1 L53 & CICE4 & CLM4 \\
NCC & NorESM1-ME & CAM4-Oslo & 2.5 x 2.5 L26 & NorESM-Ocean & 1.1 x 1.1 L53 & CICE4 & CLM4 \\
    \hline
  \end{tabular}
  \end{tiny}
  \end{center}
  \caption{Structural details of the 37 CMIP5 models included in the analysis
           of Arctic surface temperature.
           Models highlighted in red are included in the exchangeable 
           ensemble.
           Atmosphere and ocean resolution are in degrees and Lxx indicates 
           the number of vertical levels.
           Details included in this table were gathered from the metadata
           included in the model outputs and supplemented using information 
           from Table 9.A.1 of \citet{Flato2013}.\label{tab:coex_models}}
\end{sidewaystable}

In the main text we noted that not all of the models should be included in the analysis in order to satisfy the assumption of exchangeability. 
In particular, models from the same center are likely to be more similar than those from different centers. 
Therefore, only one model from each center should be included. 
Modeling centers may also share components with other groups. 
Therefore, where possible only one model using any given major component, or at least any combination of components, should be included. 

The full ensemble was thinned in order to satisfy the judgment of 
exchangeability between the model outputs. 
The ACCESS models supersede the CSIRO-Mk3.6.0 model, however all of the major components in the ACCESS models are borrowed from other models. 
Therefore, none of the models submitted by CSIRO were included. 
Two models were submitted by BCC, the model with the higher resolution atmosphere components was retained.
Three models were submitted from the combined efforts of the NSF-DOE-NCAR. 
The CESM1(CAM5) variant was selected as it includes a more recent version of the CAM atmosphere model.
The NCAR CCSM4 model has been superseded by the CESM1 model, and so was not
included.
The two NorESM1 models are also very closely related to the CESM1 model, so was excluded.
The BNU-ESM and FIO-ESM models were also excluded since they use outdated and low resolution versions of the CAM atmosphere included in the CESM1 model. 
The two models submitted from the CMCC are both based on an old atmosphere component and a very old ocean component. 
They also lack a full land surface model, therefore neither model was included. 
The CNRM-CM5 model and EC-EARTH models are very closely related, but EC-EARTH model includes more RCP4.5 runs so was retained over CNRM-CM5.
The models from NOAA-GFDL differ primarily in their ocean component.
GFDL-ESM2G uses the GOLD ocean model, while GFDL-ESM2M and GFDL-CM3 use the MOM4.1 ocean model.
However, the MOM4 ocean model is also used in the models from the BCC, so 
GFDL-ESM2M and GFDL-CM3 are excluded. 
The NASA GISS-E2-R model was retained over the GISS-E2-H for the increased number of levels in the ocean model.
The MOHC model in its HadGEM2-CC configuration has a relatively low resolution ocean component compared to most of the other models, so it is excluded in favor of the HadGEM2-ES configuration.
The model submitted by NIMR/KMA is another version of the MOHC model, and so was excluded.
The resolution of the atmospheric component of the IPSL-CM5A-LR model is also low compared to the rest of the ensemble, so it is excluded in favor of the IPSL-CM5A-MR configuration. 
Similarly, the atmospheric resolution of the MIROC models in their MIROC-ESM configuration is relatively low, so they are excluded and MIROC5 is retained.
In contrast, the MPI-ESM-MR configuration features a very high resolution ocean component compared to the rest of the ensemble. 
Therefore the MPI-ESM-LR configuration is retained instead. 
This leaves an ensemble of 89 runs from 13 models, 50 runs from the historical scenario and 39 from the RCP4.5 scenario.

\newpage
\section*{Posterior parameter estimates}

\begin{figure}
  \begin{center}
    \includegraphics[width=0.45\textwidth]{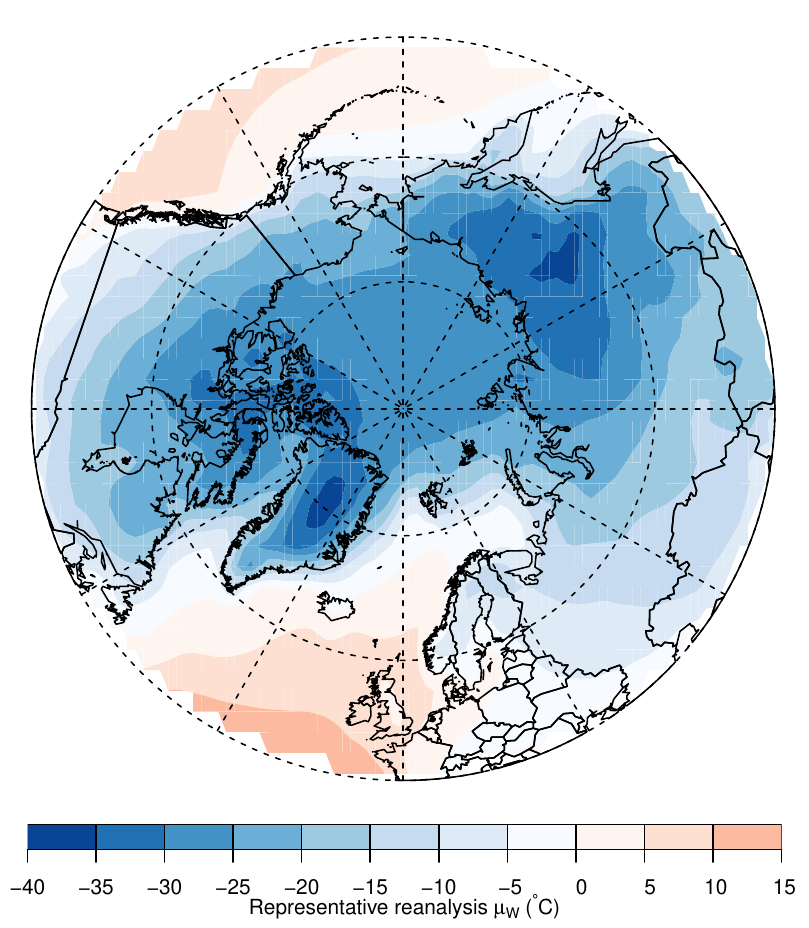}
    \includegraphics[width=0.45\textwidth]{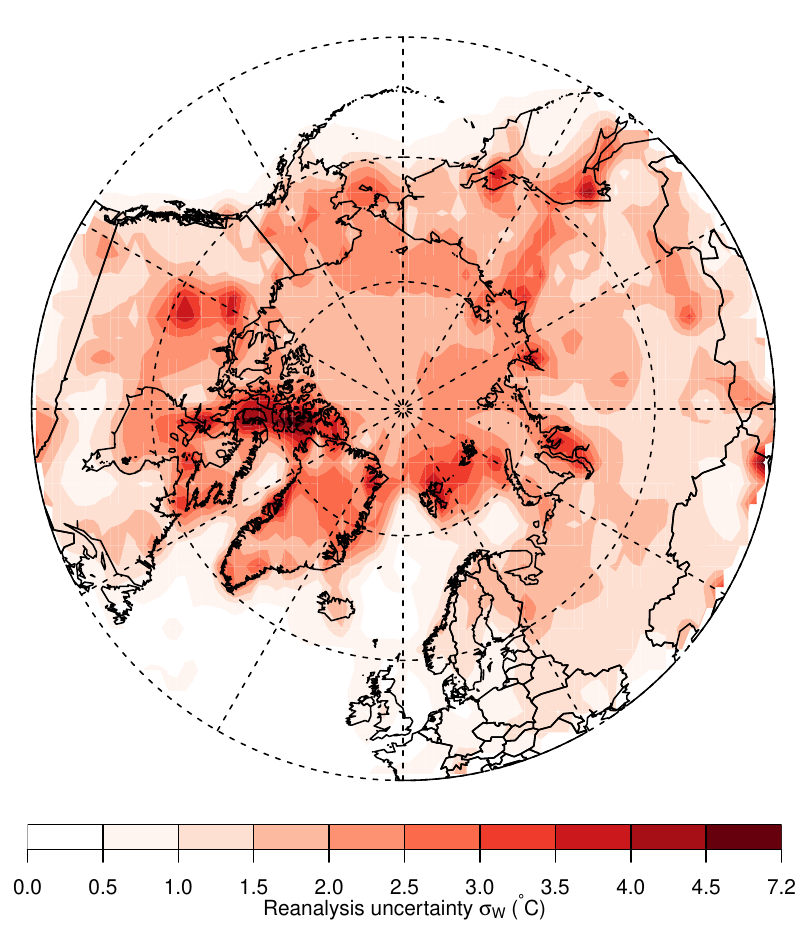}
  \end{center}
  \caption{Posterior mean estimates of the representative reanalysis $\mu_W$ and the reanalysis uncertainty $\sigma_W$.}
  \label{fig:reanalysis}
\end{figure}

The spread between the reanalyses $\sigma_W$ is greater over land than over the ocean where temperatures vary more slowly (\autoref{fig:reanalysis}).
The reanalysis uncertainty increases with latitude as the number of observing stations decreases and the terrain tends to become more mountainous (\autoref{fig:reanalysis}).
The spread between the reanalyses is particularly large around the sea ice edge.

\begin{figure}
  \begin{center}
    \includegraphics[width=0.45\textwidth]{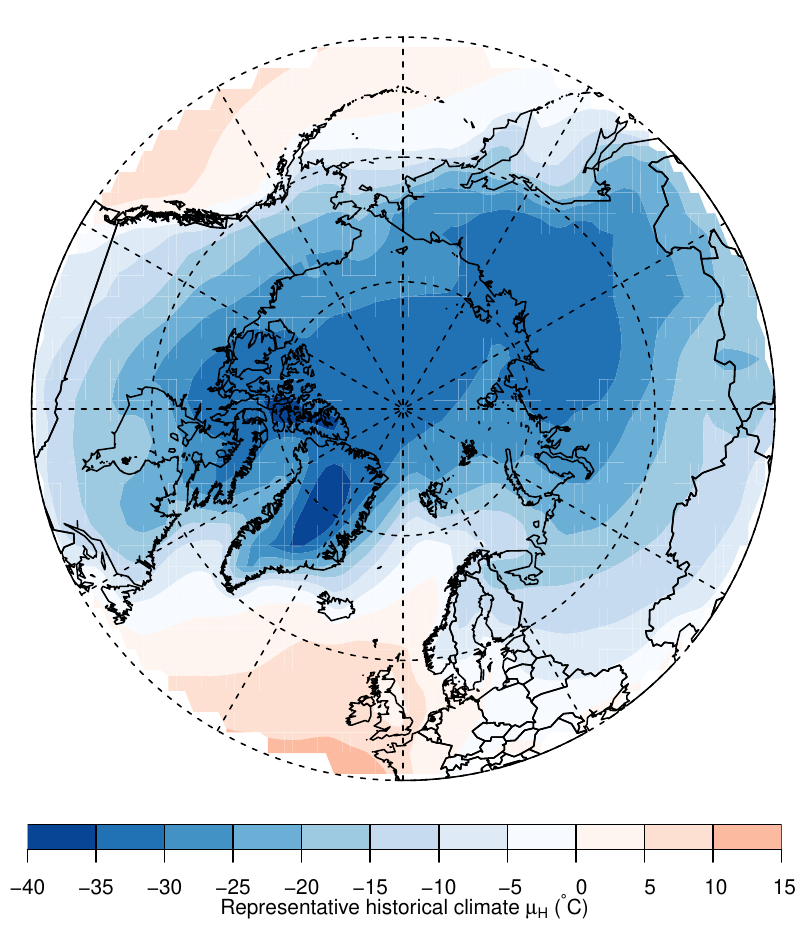}
    \includegraphics[width=0.45\textwidth]{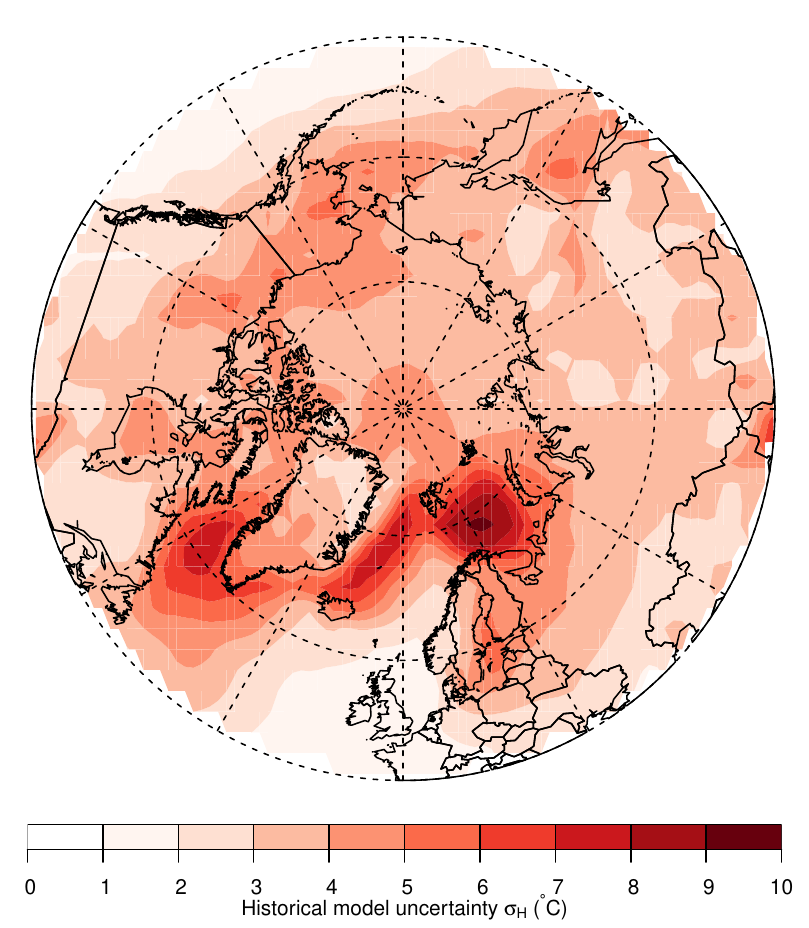}
  \end{center}
  \caption{Posterior mean estimates of the representative historical climate $\mu_H$ and the historical model uncertainty $\sigma_H$.}
  \label{fig:historical}
\end{figure}

The representative historical climate $\mu_H$ is quite similar to the representative reanalysis $\mu_W$, except over the Arctic ocean where climate models tend to be cold biased (\autoref{fig:historical}).
The historical spread between the models $\sigma_H$ is generally greater than the spread between the reanalyses $\sigma_W$ (\autoref{fig:historical}).
Like the reanalyses, the model spread tends to be greatest over mountainous regions and near the sea ice edge.

\begin{figure}
  \begin{center}
    \includegraphics[width=0.45\textwidth]{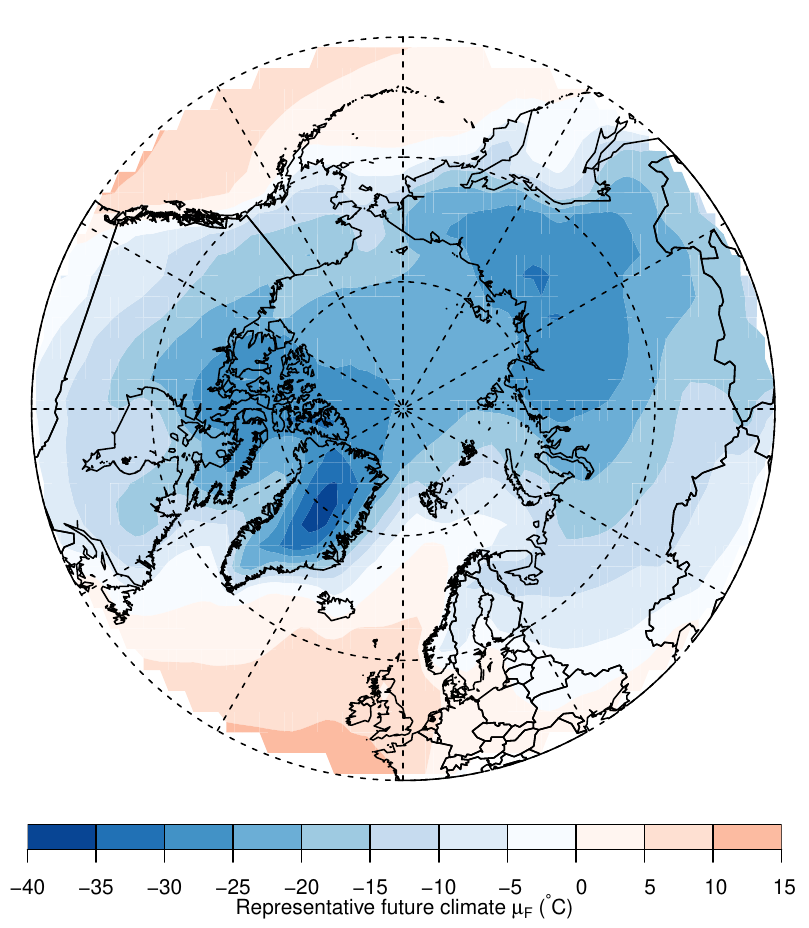}
    \includegraphics[width=0.45\textwidth]{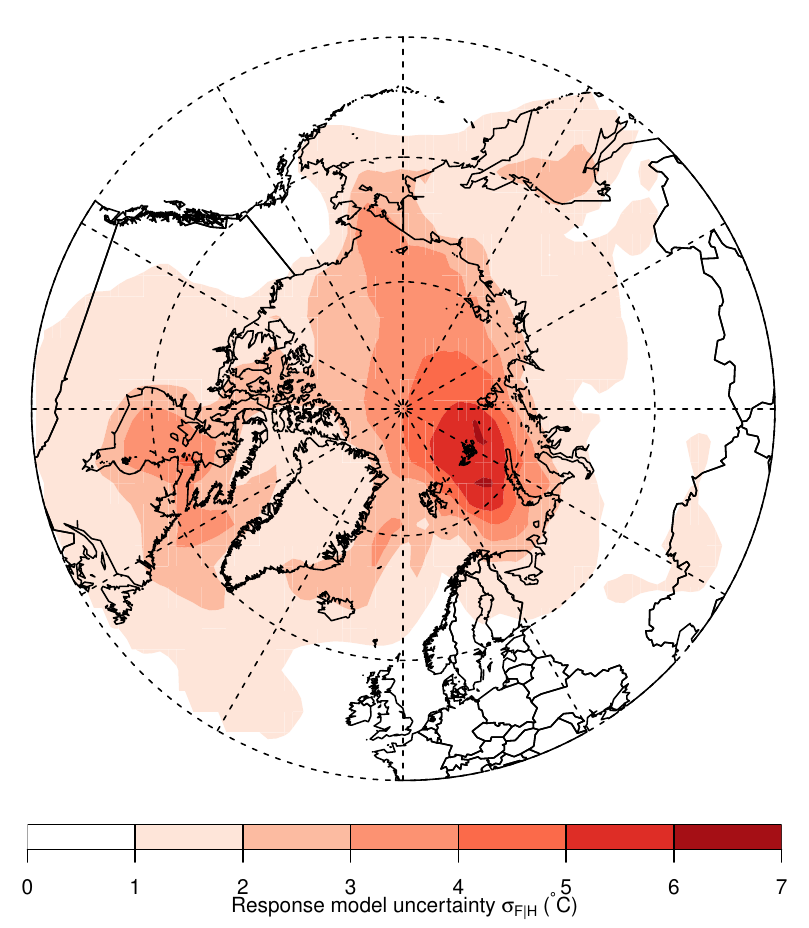}
  \end{center}
  \caption{Posterior mean estimates of the representative future climate $\mu_F$ and the model response uncertainty $\sigma_{F \mid H}$}
  \label{fig:future}
\end{figure}

The model response uncertainty $\sigma_{F \mid H}$ is greatest over the Arctic ocean, particularly to the east of Svalbard (\autoref{fig:future}).

\begin{figure}
  \begin{center}
    \includegraphics[width=0.45\textwidth]{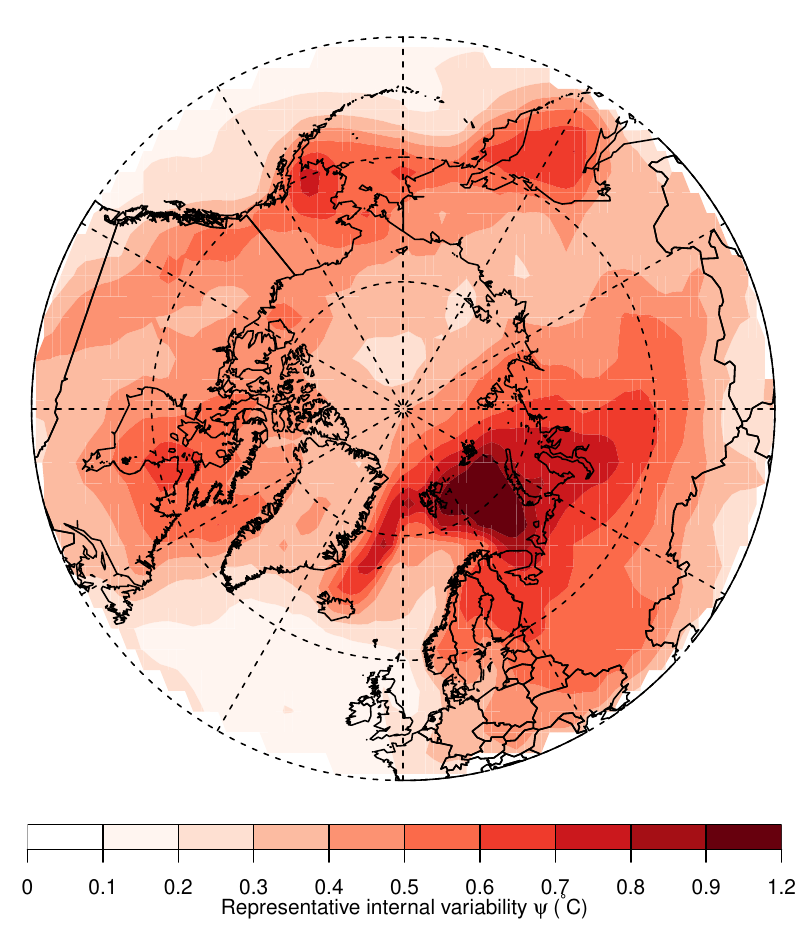}
    \includegraphics[width=0.45\textwidth]{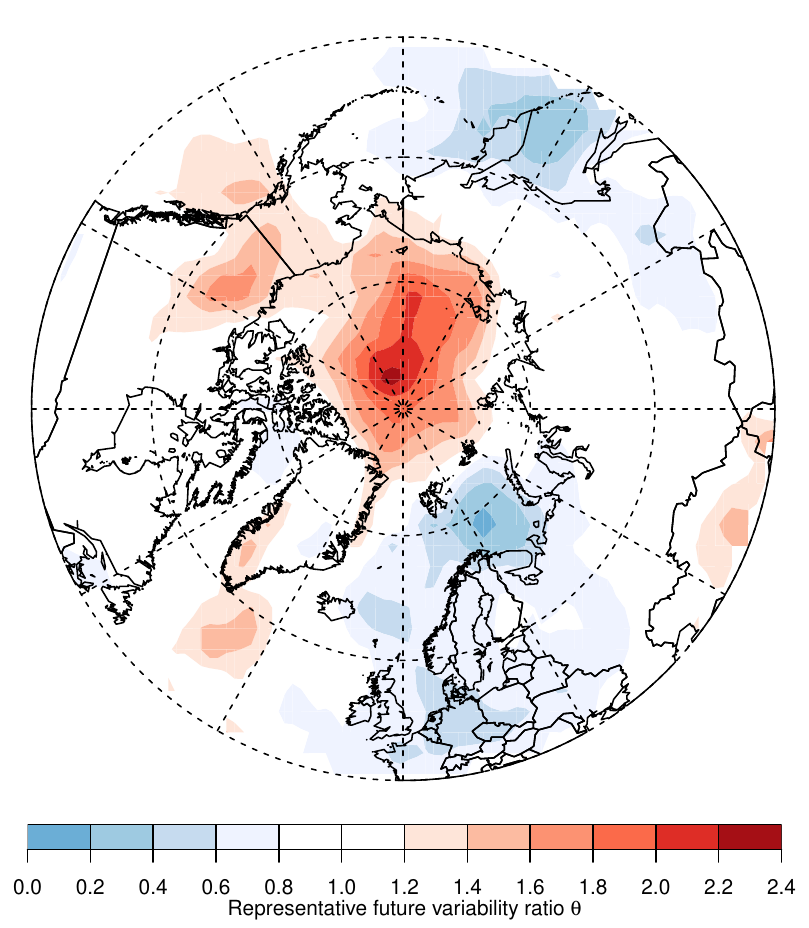}
  \end{center}
  \caption{Posterior mean estimates of the representative historical internal variability $\psi$ and change in internal variability $\theta$.}
  \label{fig:variability}
\end{figure}

Like the reanalysis uncertainty, the representative internal variability $\psi$ is greater over land than over the oceans, and highest in mountainous regions and close to the sea ice edge (\autoref{fig:variability}).
The representative change in internal variability $\theta$ is small over most of the study area (\autoref{fig:variability}).
Internal variability decreases close to the historical sea ice edge, where rising temperatures cause the ice edge to retreat and temperatures to stabilize.
The climate in the interior of the Arctic becomes more variable as rising temperatures causes seasonal melting in regions permanently covered by sea ice during the historical period.

\bibliographystyle{plainnat}
\bibliography{library}